# Astronomical Observations of Volatiles on Asteroids

**Andrew S. Rivkin**
*The Johns Hopkins University Applied Physics Laboratory*

**Humberto Campins**
*University of Central Florida*

**Joshua P. Emery**
*University of Tennessee*

**Ellen S. Howell**
*Arecibo Observatory/USRA*

**Javier Licandro**
*Instituto Astrofisica de Canarias*

**Driss Takir**
*Ithaca College and Planetary Science Institute*

**Faith Vilas**
*Planetary Science Institute*

We have long known that water and hydroxyl are important components in meteorites and asteroids. However, in the time since the publication of *Asteroids III*, evolution of astronomical instrumentation, laboratory capabilities, and theoretical models have led to great advances in our understanding of $H_2O$/OH on small bodies, and spacecraft observations of the Moon and Vesta have important implications for our interpretations of the asteroidal population. We begin this chapter with the importance of water/OH in asteroids, after which we will discuss their spectral features throughout the visible and near-infrared. We continue with an overview of the findings in meteorites and asteroids, closing with a discussion of future opportunities, the results from which we can anticipate finding in *Asteroids V*. Because this topic is of broad importance to asteroids, we also point to relevant in-depth discussions elsewhere in this volume.

# 1 Water ice sublimation and processes that create, incorporate, or deliver volatiles to the asteroid belt

## 1.1 Accretion/Solar system formation

The concept of the "snow line" (also known as "water-frost line") is often used in discussing the water inventory of small bodies in our solar system. The snow line is the heliocentric distance at which water ice is stable enough to be accreted into planetesimals. The placement of the snow line has varied in different models. A location just inside Jupiter helps explain the greater mass of the giant planets as they accrete a larger fraction of the mass in the solar nebula.  Although the location of the snow line before and during planet formation is uncertain, some studies show it may have fallen within the asteroid belt (*Lunine*, 2006). The location of the snow line in our solar system coinciding with the asteroid belt is consistent with observations of disks around other stars (e.g., *Su et al.,* 2013), where planetesimal rings tend to coincide with the snow line around those stars. However, it seems the relatively simple concept of a static snow line requires revisions in light of new results on several fronts. For instance, the snow line likely moved with time as the Sun's luminosity changed early in solar system history (e.g., *Martin and Livio,* 2012). Even though there is no agreement on the details of planetesimal formation and growth, some models favor the growth of ~100 km-scale planetesimals directly from cm-scale pieces (*Morbidelli et al.,* 2009*; Cuzzi et al.,* 2010*;* for a dissenting view see *Weidenschilling*, 2011).  Hence, the timing of those assemblies as conditions change could have a strong influence on the amount of radioactive elements present in a planetesimal's interior and on its rock-to-ice ratio, further complicating the concept of a snow line. Recent observations also point to a more complex picture. For example, the concept of an asteroid-comet continuum, where there is no clear boundary between primitive asteroids and cometary nuclei  (e.g., *Gounelle,* 2011) is gaining support and is consistent with the intermittent cometary behavior of some main belt and near-Earth asteroids (e.g., *Hsieh and Jewitt,* 2006*; Licandro et al., 2007,* 2011b*; Jewitt et al.,* 2013, Mommert et al. 2014) which will be discussed in more detail Section 1.3 below.  The chapter by *Johensen et al.* in this volume summarizes the current state of thought about asteroid formation.

1.2   **Grand Tack/Nice model delivery of outer SS material**

One of the main motivations for studying the asteroid belt is to better understand the conditions in the solar nebula and the material that formed the terrestrial planets. The distribution of spectroscopic asteroid types is not randomly mixed, but instead follows a trend with heliocentric distance (*Chapman et al.* 1975, *Gradie and Tedesco* 1982),  which has long been interpreted as a remnant property of the solar nebula. However, discoveries of extra-solar planets and other planetary systems have greatly broadened our notions of what is "normal" and "typical". Also, increasing computational capabilities have allowed ever more realistic and detailed models of accretion, and are now challenging long-held ideas of what we thought we knew about solar system formation. The asteroid belt is an important key to constraining and testing these models.

Dynamical models suggest the asteroid belt of today may be very different from when it formed. *Bottke et al.* (2012) argue that the asteroid belt once extended significantly sunward compared with today, with the Hungaria region the eroded remnant of a once-larger population. Similarly, there are arguments that Vesta formed closer to the Sun and was later transported into the main asteroid belt  (*Bottke et al.,* 2006).  More importantly for the discussion here, the "Nice Model" (*Gomes et. al.,* 2005; *Morbidelli et al.,* 2005;

*Tsiganis et al.,* 2005), which has gained wide acceptance, predicts a large influx of volatile-rich objects from the outer solar system to the inner solar system and the insertion of primitive trans-Neptunian objects into the outer belt (*Levison et al.* 2009).

The related "Grand Tack" model (*Morbidelli et al.,* 2010; *Walsh et al.,* 2011) interprets the low-albedo asteroids as originating among the giant planets, with later delivery to their current positions in the main asteroid belt by the migration of Jupiter. *Walsh et al.* (2011) and references therein describe how the original material in the asteroid belt would be disturbed by inward migration of Jupiter, leaving only the inner asteroid belt intact, scattering some of these objects outward behind it as it moves. After Saturn moves into resonance with Jupiter, both begin to migrate outward, and some of the previously scattered objects are scattered back into the asteroid belt region, along with other more volatile-rich objects (perhaps the antecedents of the C-complex asteroids) that may have formed closer to 5 AU.

This scenario solves the problem of having both high and low temperature materials side by side in the asteroid belt. However, it means that the compositional gradient with heliocentric distance is not primordial, and allows a larger amount of radial mixing than previously considered. The problem of linking meteorites to their asteroid parent bodies also becomes more complex, since the current position of the asteroids may be unrelated to where those bodies formed. However, asteroid families, collisional fragments of once intact parent bodies, are nearly all spectrally uniform, while many meteorites clearly show that differentiation has occurred. Differentiated objects can form closer to the sun, and be emplaced in the asteroid belt by Jupiter during its migration. Undifferentiated objects form farther out, and are scattered inward by Jupiter. Both can now exist side by side, but neither is necessarily representative of material that condensed at 2-4 AU.

## 1.3 Hydroxyl/water creation via space weathering

Although the term "space weathering" is usually associated with nanophase iron (or iron sulfide) embedded in glassy rims of silicate grains and the links between S-class asteroids and ordinary chondrites, the processes that form these rims (and npFe$^0$ particles) can also involve water and OH.

*Starukhina* (2001), modeling the interaction of solar wind protons with regolith, predicted that OH and its diagnostic 3-μm absorption band should always be present on airless bodies. While the quantitative results have not been borne out (for instance, Vesta has a 3-μm band depth ~2% at its deepest (*De Sanctis et al.,* 2012), compared to the ~20-70% depth predicted by Starukhina), it was one of the first treatments attempting to predict the effects of space weathering in that spectral region. Thoughts on how the spectrum of airless bodies might change near 3 μm with exposure to typical regolith processes have varied from a prediction of dehydration (*Pieters*, personal communication) to the creation of bands (e.g., *Starukhina*, 2001) to no appreciable effect (*Rivkin et al.,* 2003).

The spectral models by Starukhina followed decades of lunar-centered work by other researchers. *Arnold* (1979) investigated ice survivability on the Moon, following up the work of *Watson et al.* (1961). Arnold proposed that solar wind reduction of iron in the lunar regolith, such as is now implicated in space weathering (*Noble et al.,* 2001), could create significant amounts of water, which could then potentially make its way to cold traps in permanently shadowed craters near the lunar poles. Lunar Prospector data shows a hydrogen enhancement near the lunar poles, interpreted as due to ice (*Feldman et al.,* 2001). The independent observation of 3-μm bands on the Moon by three spacecraft (VIMS on Cassini: *Clark,* 2009, HIR-IR on the Deep Impact (DI) spacecraft: *Sunshine et al.,* 2009, the Moon Mineralogy Mapper ($M^3$) on Chandrayaan-1: *Pieters et al.*, 2009), provided convincing evidence of OH on the Moon. While interpretations are continuing to evolve, the early consensus seems to be that the band is due, at least in part, to adsorbed water and OH likely created by the interaction of solar wind protons and silicates in the lunar regolith. The presence of ice at the lunar poles is consistent with formation of water in the regolith via solar wind interactions, but other potential water sources have been identified (such as retention and migration of cometary volatiles after impacts: e.g. *Stewart et al.* 2011) and solar wind interaction is not the only possible origin for lunar water. The relative contribution and importance of any solar wind-created water to any ice deposits is not yet fully understood.

The presence of water on the lunar surface potentially reopens interpretations of asteroidal (and other small body) surfaces. However, the small bodies also provide an opportunity to better understand the lunar results. If the 3-μm band seen on the Moon is due to the implantation of solar wind protons in the lunar regolith, this absorption band should also be present on other small body surfaces, with allowances made for the increased distance from the Sun and other factors. If independent of local composition, as suggested by the high spatial resolution $M^3$ data, most asteroids should have similar effects when under similar conditions. If temperature sensitive, as suggested by the DI data, differences may be seen with solar distance. If due to photometric effects but not temperature, a well-conceived program of NEA observations should be able to separate those effects. These are still active areas of research, with a better understanding of the relationship between regolith processes and hydroxyl/water content likely over the next decade.

In addition to the possibility of solar wind protons creating water or hydroxyl, space weathering processes may have additional effects on asteroidal surfaces. *Britt et al.* (2014) argued that the micrometeorite and UV processes that create nanophase iron in silicates would, given their different mineralogies, instead create amino acids on carbonaceous bodies like the C asteroids. They suggest that space weathering, when viewed from a thermodynamic standpoint, involves decomposition of rock-forming minerals with results depending on specific composition. In their view, this decomposition is only the first stage of space weathering, with a second stage using the results of that decomposition as catalysts. For the volatile-rich carbonaceous chondrites and their parent asteroids, not only silicates but water, carbon monoxide, and other volatiles can interact via Fischer-Tropsch reactions to create ever-longer carbon chains and kerogen-like insoluble organic matter, as is seen in carbonaceous chondrites. This line of research is still in its infancy, and the observational consequences and predictions are not yet fully understood, particularly in the context of water. Previous work on space weathering of carbonaceous chondrites and/or low-albedo asteroids has focused on changes in spectral slope rather than how it may alter absorptions associated with hydrated minerals.

## 1.4 Delivery of exogenic hydrated material

It has been proposed that dark material identified on the surface of asteroid 4 Vesta by NASA's Dawn spacecraft was delivered by low-velocity impacts with carbonaceous chondritic material (*Reddy et al.,* 2012). Several lines of evidence support this interpretation, including elemental composition (*Prettyman et al.,* 2012), near-infrared spectra (*McCord et al.,* 2012), geomorphology (*Denevi et al.,* 2012), and analogy with HED meteorites, especially Howardites (*Reddy et al.,* 2012 and references therein). These findings confirm reports of a weak absorption near 3 µm in ground-based spectroscopy of Vesta by *Hasegawa et al.* (2003) and *Rivkin et al.* (2006), who suggested that impacts with carbonaceous objects could have added to the original igneous surface composition.

Visible images from the Dawn spacecraft place most of the dark material in the oldest regions of Vesta's surface. Initially two sources for this dark material were considered, ancient volcanic activity or exogenic material. However, convergence among the different observational constraints points to an exogenic origin for the dark material; Further details appear in the chapter by *Russell et al.* in this volume.

The same exogenic source of hydrated carbonaceous chondritic material on Vesta is proposed by *Shepard et al.* (2015) to explain a 3-µm absorption feature, often attributed to the presence of hydrated minerals (*Rivkin et al,* 2000), on a number of M-class asteroids with the high radar reflectivity suggestive of a high metal content. In a related result, *Landsman et al.* (2013) reported the detection of 3-µm absorption features on five M-class asteroids recently observed at these wavelengths. *Landsman et al.* (2013) also report considerable diversity in the shape of the hydrated mineral feature in their M-class asteroid sample. However, the M class in general contains a great deal of variety, and objects with low radar reflectivity and 3-µm bands also appear in this class. It seems likely that at least some M asteroids do not have high metal content. Further discussion of the M-class asteroids is found in Section 3.3.4.

Interestingly, exogenic processes have also been invoked by *Bottke et al.* (2010, 2013), who argued that impact debris from the irregular satellites of the Jovian planets should be driven onto the regular satellites of those systems, and could have accumulated to a depth of 10s or even 100s of meters on the Galilean satellites. The irregular satellites are further discussed below.

## 2 Discussion of spectral features on asteroids and meteorites

While two sample returns to low-albedo asteroids are slated for the next decade (Section 6.2), remote sensing will by necessity remain the primary method for detecting and characterizing the mineralogies of large numbers of small bodies. Three wavelength regions are primarily used for this purpose, each with their own strengths and limitations: the visible and near infrared spectral region (~0.4-0.8 µm), the 3-µm region, and the mid-IR (5-13 µm). An overview table of the most important absorption bands is presented in Table 1. Other wavelengths are used in cometary studies, for instance, but have not been used at asteroids due to an expectation of very low contrast and very meager gas production rates from asteroids. Other techniques, such as ground-based radar or sub-mm observations may give indirect evidence for the presence of hydrated minerals (*Hanson et al. 2006*), but our understanding of the relationship between radar albedo and hydrated minerals is incomplete. The meteorites have given us most of our information about the specific minerals present in the asteroids and the aqueous alteration processes that occurred early in solar system history. We discuss the spectral properties of the hydrated minerals and aqueous alteration products below, but details of the minerals in and processes experienced by the meteorites themselves, and the conditions under which they were created and operated are discussed in detail in the chapter by Krot et al. in this volume.

We note that there are additional ways to detect water/OH and other volatiles, such as neutron spectrometers like the one carried by Dawn, or via sounding with radar instruments like the SHARAD instrument on MRO (*Seu et al. 2004*). These may be the only way to directly detect volatiles on some objects, particularly those for which surface volatiles are unstable. We acknowledge that such measurements are particularly valuable, but are out of the scope of discussion in a chapter devoted to "astronomical observations".

## 2.1 Visible/Near-IR

An absorption band centered near 0.7-µm is seen in asteroids as well as the CM meteorites. The absorption has classically been attributed to $Fe^{2+}$-$Fe^{3+}$ intervalence charge transfer, associated with phyllosilicates but not diagnostic of them (*Vilas and Gaffey*, 1989; *Vilas*, 1994; *King and Clark*, 1997). *Cloutis et al.* (2011a) attributes bands near 0.7 µm to saponite group phyllosilicates (band centers 0.59-0.67 µm) and "mixed valence Fe-bearing serpentine group phyllosilicates" (band centers 0.70-0.75 µm). Unlike the Ch asteroids, for which the 0.7-µm band is a defining characteristic (Section 4.3.1), some CM meteorites have no absorption band at these wavelengths. In addition, while a few thermally metamorphosed, non-CM meteorites also have the band (B-7904: C2-ung; Y-86029:CI1), such samples discussed in *Cloutis et al.* (2012) can be grouped with the CM meteorites for our purposes.

Combination bands involving the OH fundamental stretching mode appear in the 2.2-2.4 µm region, generally in pairs (*Hunt,* 1977). These bands are commonly used to identify phyllosilicates on the Earth and Mars, with the band positions providing the Al and/or Mg composition of the minerals. The presence of opaques strongly masks these bands (*Clark,* 1983), and there is only "occasional evidence" for them in spectra of CM chondrites (*Cloutis et al.*, 2011a) and they are only weakly found in "a few" CI chondrites (*Cloutis et al.*, 2011b). It is not surprising, then, that telescopic spectra of asteroids have not shown conclusive evidence of these bands. However, their presence in the meteorite data suggests that sufficiently high-quality data could be capable of capturing the bands, providing an additional way of measuring the composition of low-albedo asteroids.

Two additional near-IR bands of possible interest are at 1.4 and 1.9 µm. These are overtones of longer-wavelength bands: the 1.4-µm absorption is the first overtone of the OH band at ~2.7-2.8 µm (see below), while the 1.9-µm band is due to a combination of an $H_2O$ bending mode and the OH stretching modes (*Clark,* 1999). As with the phyllosilicate bands above, these absorptions are of great use in terrestrial studies, but are much more difficult to use in asteroidal studies. In addition to their susceptibility to masking by opaque components, their association with water *per se* makes them difficult to observe through the Earth's atmosphere and in laboratory measurements at ambient conditions (see, for instance, *Bishop et al.,* 1995 for spectral changes between Earth ambient and Mars ambient conditions). Because of these limitations, the 1.4 and 1.9-µm bands are unlikely to be utilized for observations of asteroidal volatiles in the near future.

There have also been investigations at wavelengths shorter than 0.5 µm, most notably focusing on an absorption at 0.43 µm that *Vilas et al.* (1993) attributed to ferric iron in aqueously altered minerals and the *U-B* color index (an indication of oxidized iron), which *Feierberg et al.* (1985) suggested was correlated to 3-µm band depth in a sample of 14 low-albedo asteroids. *Vilas* (1995) found no correlation between *U-B* color and 3-µm absorption in higher-albedo objects, suggesting little oxidized iron on their surfaces. *Rivkin* (2012) found that the *u'–g'* color in the Sloan Digital Sky Survey (SDSS) did not differ between the C-complex asteroids with and without a 0.7-µm band, suggesting it was not diagnostic for hydrated minerals. While the rise of CCDs has led to relatively few UV observations of asteroids in the past decade, the UV capabilities of spacecraft like Rosetta (*Coradini et al,* 1998, *Stern et al.,* 2011), New Horizons (*Stern et al.,* 2008), and Dawn (*De Sanctis et al.,* 2012) may spur more small-bodies work in this wavelength region.

2.2  **3-µm Region**
Water and hydroxyl both have strong absorptions in the 3-µm region. Hydroxyl has a band position that varies depending on specific composition but is generally expected in the 2.7-2.8 µm region (*Clark et al.,* 1990). However, the band's position can also vary with other factors: *Farmer* (1974) cites positions from 2.67-2.94 µm in layer silicates and *Ryskin* (1974) finds a band position at wavelengths as long as 3.45 µm in some materials. The 3-µm region also hosts other volatile species: methane and other organic materials have absorptions in the 3.3-3.4 µm region, and while ammonia is not expected on asteroidal surfaces, the ammonium ($NH^{4+}$) ion has a band center near 3.1 µm and was interpreted as present in Ceres' spectrum (*King et al.,* 1992). Several studies have established the carbonaceous chondrite meteorites in our collections share particular band shapes (Figure 1), with a strong absorption edge near 2.7 µm due to hydroxyl followed by a relatively linear return to continuum behavior as wavelengths increase. Earlier studies (*Hiroi et al.,* 1996; *Jones,* 1989) unavoidably retained some telluric water in their laboratory spectra, but similar spectral behavior is seen in samples in which telluric water has been removed (*Beck et al.,* 2010; *Takir et al.,* 2013, Section 3 below). However, in those cases where terrestrial adsorbed water is still present, the absorption band is deeper and wider. Studies have shown that the band minimum in the 2.7-2.8 µm region in carbonaceous chondrites is indicative of phyllosilicate composition and degree of aqueous alteration (*Sato et al.,* 1997; *Osawa et al.,* 2005; *Beck et al.,* 2010; *Takir et al.,* 2013).

The benefits derived from observing absorptions due to water and OH *per se* rather than indirectly via associated absorptions are partially offset due to the difficulties associated with observing at these wavelengths. Transmission through the Earth's atmosphere is closely related to the spectrum of water vapor, and the wavelengths with the strongest absorptions in hydrated and hydroxylated minerals typically have low atmospheric transmission. Indeed, most researchers will simply omit data obtained in the 2.5-2.8 µm spectral region as compromised beyond recovery. Furthermore, detailed below, laboratory measurements of materials are not immune to interference from terrestrial water. Early researchers found that atmospheric water quickly affected samples measured in ambient conditions, with atmospheric water readsorbing on samples between the time they were removed from a dry nitrogen atmosphere and the time they were measured in the spectrometer, and some adsorbed water remained even under a strong vacuum at room temperatures (*Clark et al.,* 1993).

Comparing meteorite and asteroid reflectance spectra in the 3 µm region can therefore be challenging because meteorite spectra have usually been acquired under ambient terrestrial conditions where atmospheric water is dominant (see above), but advances in laboratory spectroscopy allow samples to be heated and/or put into vacuum to remove terrestrial water and spectra to be collected without exposing the sample to ambient conditions.

In addition to the measures to improve laboratory data, techniques for asteroid observations in this wavelength region have also been developed. The absorption bands seen in minerals in the 3-µm spectral region are broader than those seen in the atmosphere and with care and effort the difficulties can be limited or minimized. The vast majority of observations in this wavelength region are obtained at Mauna Kea, with precipitable water vapor values commonly less than 10% of the mean U.S value of 20 mm (*Shands,* 1949), and reduction techniques have been developed to fit and remove atmospheric lines from object and standard spectra, minimizing the effects of airmass mismatch and increasing the number of calibration stars that can be fuitfully used for a given object (*Volquardsen et al.,* 2007).

Water *per se* has a different band shape and band position than hydroxyl. Water ice has three separate bands due to molecular vibrations in the 3-µm region, at ~3.2, ~3.1, and ~3.0 µm, with shifts in center depending on whether the ice is crystalline or amorphous, and as a function of temperature (*Mastrapa et al.,* 2009). The strength and proximity of these bands results in a broad "3-µm band" due to their overlap. The strength of the water band is such that thicknesses of more than ~100 nm result in band saturation. This is inevitably the case on icy satellites, but ices of this thickness or less have been reported on asteroids (*Rivkin and Emery* 2010, *Campins et al.* 2010).

In both the 3-µm region and the mid-IR (section 2.3), contributions by thermal emission from the asteroid to the measured spectrum can be important—indeed, by 4-5 µm thermal emission dominates over reflected light throughout the main asteroid belt. In order to make useful comparisons to laboratory data, this thermal flux must be modeled and removed. The techniques for doing so typically use the Standard Thermal Model (STM) or Near Earth Asteroid Thermal Model (NEATM). While beyond the scope of this chapter, details of these models can be found in *Lebofsky and Spencer* (1989) and *Harris et al.* (2002), and *Delbo et al.* in this volume.

Thermal flux can also, in some circumstances, "fill in" an absorption band. Because these circumstances can arise in the 3-µm spectral region, of particular interest for hydrated asteroid studies, here we discuss those circumstances and the consequences for astronomical observations following the logic presented by *Rivkin et al.* (2013a) and *Clark* (1979). In an absorption band, the emissivity increases relative to that outside the band because the reflectance decreases in the band. The increased emissivity leads to an increase in thermal flux inside the band, potentially enough to "fill in" the absorption at sufficiently high temperatures. It can easily be shown that halving the band depth (or doubling it) would similarly halve (or double) the change in emissivity, leading to the same level of fill-in at a given temperature regardless of band depth. Roughly speaking, to erase an absorption band, the thermal flux times the continuum reflectance at the wavelength of interest needs to be about the reflected flux. This condition does not occur in the 3-µm region for main-belt asteroids, but can be reached by 3.5 µm or shortward in NEOs (*Rivkin et al.,* 2013b), and partial filling can occur at shorter wavelengths.

This treatment of thermal fill-in only formally holds in situations where Kirkhoff's Law is valid. The bi-directional nature of the reflectance data collected for asteroids vs. the hemispherical-directional nature of the emission data collected means that Kirchoff's Law does not fully hold, and Hapke modeling of organic and carbon dioxide bands suggests fill-in may occur for thermal fluxes ~30-50% of what the rule of thumb above would suggest. It is likely that the OSIRIS-REx mission will see band depths appear to change with time of day as surfaces on Bennu heat and cool, and it is possible that some observations of organic materials may be difficult or impossible at some times of day due to this effect. Such observations, however, will be useful for more rigorously understanding the conditions under which thermal fill-in is important.

## 2.3 Mid-IR Region

The mid-infrared (5-50 µm) spectral region has not been widely used to measure asteroid mineralogy, including hydration state. The main reasons for this neglect are: 1) spectral contrast is generally quite low and 2) the high flux and rapid variability of thermal emission from earth's atmosphere has made it difficult to obtain spectra of sufficient sensitivity. Nevertheless, several mid-IR spectral features offer the opportunity for evaluating hydration state, and space platforms and large ground-based telescopes have produced very high quality mid-infrared data (see chapter by *Reddy et al.* in this volume).

The primary feature for direct detection of hydration in the mid-IR is the H-O-H (or X-O-H) bending vibration near 6.25 μm. For typical planetary materials, the strongest features in this wavelength range are the Si-O stretch and bend vibrational modes of silicates near 8–12 μm and 15–25 μm, respectively. Since bond strengths (and therefore vibrations) are sensitive to molecular structure, silicates express a rich diversity of mid-IR spectra that are diagnostic not only of silicate class, but also of compositions. Phyllosilicates, particularly those typically found in carbonaceous meteorites, tend to have a single-peaked absorption in the Si-O stretch region near 10 μm. *Beck et al.* (2014), using transmission spectra, find a consistent anti-correlation between the location of this absorption and the presence of an olivine peak near 11.2 μm for CI and CM meteorites, which they interpret as an indication of degree of hydration. The 6.25-μm feature is not generally present in their spectra, but the transmission pathlengths may have been insufficient. It is not clear whether reflectance or emittance spectra would show such a trend, but mid-IR reflectance spectra of several CI and CM meteorites published by *Salisbury et al.* (1991) show indications of similar behavior. The reflectance spectra also contain a ~6.15-μm feature, particularly in the more heavily hydrated meteorites, but the spectra were measured in air, so contamination by adsorbed $H_2O$ is likely. Other phases associated with aqueous alteration (e.g., carbonates) also have spectral features in the mid-IR (e.g., *Clark et al.,* 2007).

Only a few applications of telescopic mid-IR data to identify hydration on asteroids have been published. *Milliken and Rivkin* (2009) show that a 6–14 μm spectrum of Ceres obtained from the Kuiper Airborne Observatory by *Cohen et al.* (1998) is well-matched by linear-mixing models that were developed for the 2–4 μm spectral region and that include brucite ($MgOH_2$) and cronstedtite (Fe-rich serpentine). *Vernazza et al.* (2013) note that 368 Haidea is the only asteroid in the Spitzer Space Telescope (SST) spectral database (of 87 asteroids searched) that shows a single emissivity peak in the Si-O stretch region. The spectrum of Haidea is well-matched in the visible–NIR and the mid-IR by the Tagish Lake meteorite, which is a heavily hydrated meteorite (*Zolensky et al.,* 2002), from which Vernazza et al. conclude that Haidea too is hydrated. SST spectra of other D-type asteroids (e.g., *Emery et al.*, 2006) show clear olivine peaks, quite distinct from Haidea and Tagish Lake, supporting the view that D-type asteroids are composed of anhydrous, primitive material and suggesting that both Tagish Lake and Haidea are somewhat anomalous. Using laboratory spectra of CM meteorites and SST spectra of three C-type asteroids, *McAdam et al.* (2013) suggest that some sub-features at ~12-14 μm in phyllosilicates may also be diagnostic of hydration. These are within the broader "transparency feature" found at those wavelengths, where the emissivity is at its lowest (see the *Reddy et al.* chapter in this volume for a fuller discussion of mid-IR spectral features).

With no active mid-IR space telescopes, with many of the mid-IR spectrometers on large ground-based telescopes not offered in recent calls, and with SOFIA's future uncertain, the immediate future for exploring hydration in the mid-IR might seem challenging.

However, CanariCam, a mid-IR camera/spectrometer on the recently built 10-m Gran Telescopio Canarias (GTC) is demonstrating that high sensitivity mid-IR spectra of asteroids can be obtained from the ground (*Licandro,*et al. 2014). The James Webb Space Telescope (JWST), with spectral capability from 1–28 μm (http://jwst.nasa.gov/about.html), should be an excellent facility for studying hydration on asteroids. And lastly, near- and mid-IR instrument payloads are becoming more common on proposed and in-development spacecraft missions, including the OVIRS and OTES instruments on NASA's OSIRIS-REx mission, which will map the presumably hydrous (e.g., *Clark et al.,* 2011) asteroid Bennu from 0.4–50 μm down to spatial scales of ~24 cm. These and other future opportunities are further discussed in Section 6.

# 3 Results and Interpretations for asteroids and meteorites

The spectral analysis of water and OH in meteorites has taken several large strides in the past decade. A series of papers by *Cloutis et al.* focused on the visible and near-IR spectra of carbonaceous chondrites. Increased laboratory capabilities, including the ability to drive off terrestrial water and take spectra without breaking vacuum, have led to a much better sense of the spectral properties of hydroxyl in carbonaceous chondrites (Section 2.2 and below), and mapped out OH band centers with mineralogy. Work by *Milliken et al.* (2007) in the context of Martian studies provides a potential way to remotely measure water in meteorites through calculation of spectral parameters such as band depth and integrated band depth, integrated band area, mean and normalized optical path length, and effective single-particle absorption thickness, though to this point these have not been applied to estimating hydroxyl amounts, nor been used on carbonaceous chondrites in the published literature.

CM (Meghei-like) and CI (Ivuna-like) carbonaceous chondrite meteorites are widely thought to be the possible analogs for dark and primitive asteroids (e.g., C-, D-, G-, K-, F-, and B-types in the Tholen taxonomy, C-complex in the Bus taxonomy: e.g., *Burbine et al.,* 2002). These meteorites experienced varying degrees of fluid-assisted alteration on their parent bodies (*McSween,* 1979). Spectroscopic, mineralogic, oxygen and hydrogen isotopic, and textural analyses of CM and CI carbonaceous chondrites have revealed that these meteorites contain aqueously altered materials (*Takir et al.,* 2013; *Hiroi et al.,* 1996; *Beck et al.,* 2010; *Zolensky et al.,* 1993; *Clayton and Mayeda,* 1999; *Lee,* 1993, and others). Laboratory analyses of carbonaceous chondrites combined with astronomical observations of hydrated asteroids can lead to important clues about abundance and distribution of $H_2O$ in the early Solar System.

*Takir et al.* (2013) measured meteorite spectra under dry (elevated temperatures) and vacuum ($10^{-8}$ to $10^{-7}$ torr) conditions to minimize the effect of adsorbed water and mimic the asteroid-like environment. Beck et al. (2010) previously measured carbonaceous chondrite transmission spectra; however, absorption features in transmission do not always have the same shape or position as they do when measured in reflectance, due to the scattering of reflected illumination by granular particles. *Takir et al.* (2013) also investigated the degree of hydration in CM and CI chondrites, using the previously defined alteration scales of *Browning et al.* (1996)(Mineralogical Alteration Index) and

*Rubin et al.* (2007) (Petrological Subtype). *Takir et al.* identified three spectral groups of CM chondrites (plus the CI chondrite Ivuna), using the 3-μm band center and shape of Infrared (IR) reflectance spectra (Figure 1). The diversity in the 3-μm band demonstrates that distinct parent body aqueous alteration environments experienced by carbonaceous chondrites can be distinguished using reflectance spectroscopy. The first group (Group 1) shows the lowest degree of aqueous alteration and is characterized by 3-μm band centers at longer wavelengths than other groups. Group 1 is consistent with the occurrence of Fe-rich serpentine (cronstdetite). Group 3 that exhibits the highest degree of aqueous alteration is characterized by 3-μm band centers at shorter wavelengths and is consistent with Mg-rich serpentine (antigorite). The third group, Group 2, is intermediate between Groups 1 and 3 and more likely includes both types of serpentine minerals.

Along with the advances in meteorite spectroscopy, observations of asteroids have flowered in the past decade. The majority of telescopic observations in the 3-μm region have been obtained at the IRTF using the SpeX instrument, but observations at other wavelengths have been made using other telescopes across the world, with orbiting telescopes, and *in situ* orbiting Vesta and flying past Lutetia. Asteroid researchers have also made use of datasets acquired for astrophysical projects such as the SDSS. The use of all these observations has led to a fuller characterization of several individual objects and an understanding of the hydrated mineralogy present in the asteroid belt and beyond.

## 3.1 Correlation of 0.7-μm Band and 3-μm Band:

A growing database of asteroidal spectra available at both 0.7 and 3 μm has been accumulating in order to understand the relationship between these bands. *Vilas* (1994) showed that the 0.7-μm band is present when the 3-μm band is seen in ~85% of a sample of 31 CCD reflectance spectra and IR multicolor photometry observations of low-albedo asteroids, and applied this correlation to the larger ECAS photometry database to identify potentially aqueously-altered asteroids. While this is good evidence that the bands are related, it is also important to understand why the bands sometimes are not correlated. *Howell et al.* (2011) show that, upon closer examination using simultaneous observations, the 3-μm band is often variable with rotation of an asteroid, and that whenever the 0.7-μm band is present, the 3-μm band is also seen. However, the 3-μm band can be present without the 0.7-μm band, and indeed is seen in about half of the objects observed at 3-μm for which the 0.7 μm band is absent. This makes the 0.7-μm absorption a reliable proxy for hydration, but the 0.7-μm absorption is limited to only establishing a lower bound on the total number of hydrated objects. Table 2 shows the correlation of the 156 objects in the literature that have been observed at both wavelengths.

## 3.2 3-μm Band Shapes:

Using high-quality near-infrared (0.7-4.0 μm) spectra of 35 outer Main Belt asteroids (2.5 < a < 4.0 AU), *Takir and Emery* (2012) identified four 3-μm spectral and orbital groups, each of which is presumably linked to distinct surface mineralogy (Figure 4). Table 3 shows these groups, examples of each, alternate names, and compositional interpretation. The sharp group (or Pallas types) exhibits a characteristically sharp 3-μm feature, reflectance decreasing with decreasing wavelength into the 2.5-2.85 μm spectral region, attributed to OH-stretching in hydrated minerals (e.g., phyllosilicates) (*Rivkin et al.,* 2002b). The majority of asteroids in this group are concentrated in the 2.5 < a < 3.3 AU region. The second group, the rounded group (or Themis types), exhibits a rounded 3-μm band (reflectance increases with decreasing wavelength shortward of ~3.07 μm), attributed to $H_2O$ ice (e.g., *Rivkin and Emery,* 2010). Asteroids in this group are located in 3.4 < a < 4.0 AU region. The third group, 1 Ceres-like group, is located in the 2.5-3.3 AU region and characterized by a narrow 3-μm band center at ~3.05 μm superposed on a much wider absorption from 2.8 to 3.7 μm that is consistent with brucite (*Milliken and Rivkin*, 2009). The fourth group, the 52 Europa-like group (grouped into the Themis types by *Rivkin et al.*) exhibits a 3-μm band centered around 3.15 μm with longer wavelength band minimum and steeper rise on the long-wavelength edge of the absorption.
On the basis of the 3-μm band shape and center, *Takir et al.* (2014) found that the sharp asteroids possibly have similar phyllosilcate mineralogy as CM Group 2 and Ivuna, suggesting that these asteroids and meteorites experienced similar aqueous alteration processes. These results suggest that CM and CI chondrites are possibly the meteorite analogs for the sharp group (Figure 3). The authors found Ivuna to have a spectrum consistent with lizardite and chrysotile. No meteorite match was found either for the rounded group, Ceres-like group, or Europa-like group.

Differences within each of these groups are still to be fully understood. For instance, the relationship between the Europa, Ceres, and rounded groups is not entirely clear—the Ceres types are interpreted very differently from the rounded group, but there may be objects that appear spectrally intermediate between the groups.  It is apparent that there is variation within the sharp group, with some structure visible within the main absorption on some objects. It is also not certain whether the same object can vary between different band shapes on its surface, for example 704 Interamnia is classified differently in *Takir et al* (2012) and *Rivkin et al.* (2012). Further work is necessary to better establish the compositions found in these groups: *Beck et al.* (2010) proposed goethite as an alternate to ice frost in the Themis-type asteroids based in part on the thermodynamic difficulty of keeping frost on these asteroid surfaces. Nonetheless, while other closely related minerals have been found in both meteorite and asteroid spectra, extraterrestrial goethite has never been identified within the meteorite inventory. *Jewitt* (2012), considering ice vs. goethite, preferred ice as a simpler interpretation for this reason. Also, goethite is a poor match for the spectrum of 24 Themis at wavelengths shortward of 3 μm (*Pinilla-Alonso et al.,* 2011). The interpretation of brucite and carbonates on Ceres will be tested and honed by Dawn's visit to that object, with consequences for the interpretation of other Ceres types.

Additional groups beyond those identified in Takir et al. and mentioned above appear to be present. *Rivkin et al.* (2011) found the band shape on Lutetia to be unlike Pallas, Ceres, or Themis (Figure 4) and proposed a fit to goethite, inspired by *Beck et al.* (2011). *Yang et al.* (2009) observed a flat-floored 3-μm band on Comet Holmes due to ice grains, which may theoretically be present on portions of asteroid surfaces, though such a spectrum has not yet been found on asteroids. Finally, some objects appear to have no absorption band in the 3-µm region. In some cases, like the Trojan asteroids, this may be because only relatively poor-quality data is available and significant absorptions may be hidden in the uncertainties. However, resolved observations of comets 9P/Tempel 1 and 103P/Hartley 2 by *Groussin et al.* (2013) showed spectra were very well fit by a sum of reflected and thermally emitted flux, showing no 3-µm absorption within ~2% (although $H_2O$ emission was seen in some cases).

Preliminary results from the LXD-mode Main-belt and NEO Observing Program (LMNOP, *Rivkin et al.* 2012, 2014b) are consistent with Takir et al., with some additional insights available from a larger sample size. The different 3-µm types can be found in most of the C-complex component classes, other than the Ch class as discussed below. When the Ch class is excluded, Pallas/sharp types make up 44% of the 48 C-complex asteroids in the LMNOP for which band shapes can be assigned, while Themis types make up 42%. This is similar to the (smaller) X-complex sample of 13, including seven Pallas types and five Themis types. The four Ceres types in the LMNOP sample are all found in the C complex.

The sample of low-albedo (but non-Ch) asteroids shows a correlation between 3-µm band shape and size as well as with solar distance: Pallas types make up 27% of the H<7 sample (corresponding roughly to D > 200 km), but 69% of the H<8.5 sample (D < 100 km). The Themis types start at ~36% of the largest-sized sample, peaking at 55% of the 150-200 km sample before declining to 40% of the H < 8.5 group. The Ceres types are found only in the H < 7 group. Turning to solar distance (and again excluding the Ch asteroids) shows the inner belt (a < 2.82 AU) dominated by the Pallas types (63%) while in the outer belt (a > 2.82 AU) the Themis types are more numerous, outnumbering the Pallas types 48%-33%.

The observation that Ch asteroids are overwhelmingly (perhaps unanimously) Pallas-type objects means that the former can be used as a proxy for the latter, even if only providing a lower limit on their numbers. The work of Rivkin (2012) using the SDSS to study the 0.7-µm band is consistent with the LMNOP findings and is discussed below.

## 3.3 Connections with Taxonomy

### 3.3.1 C-complex Asteroids:
The majority of work on hydrated asteroids has focused on the C-complex asteroids, a group that comprises the B, C, Cb, Cg, Ch, and Cgh classes in the Bus taxonomy (*Bus and Binzel,* 2002a). These asteroids, as described elsewhere in this volume, have low albedos and are generally associated with the carbonaceous chondrite meteorites. Most of the largest bodies in the asteroid belt are C-complex asteroids, including Ceres, Pallas, Hygiea, Interamnia, and Europa.

In the Bus asteroid taxonomy, the Cgh and Ch asteroid classes are defined as those with 0.7-µm bands, with behavior shortward of 0.55 µm discriminating between the Cgh and Ch classes. The extension of the Bus taxonomy to 2.5 µm by *DeMeo et al.* (2009)

maintained those two classes, with similar definitions. The Cgh class is relatively rare, with Ch asteroids in the SMASS dataset outnumbering them 9:1. While the shared characteristic of these classes is the 0.7-μm band, the exact band minimum varies from roughly 0.687-0.740 μm in a sample of Ch asteroids studied by *Rivkin et al* (2013). *Vilas et al.* (1994) noted the difference in 0.7-μm band center between the C-class asteroids and the CM2 carbonaceous chondrites where it is nearer to 0.72 - 0.73 μm, and suggested that this difference could be due to a shift in temperature between asteroid surfaces and laboratory reflectance studies, but found no laboratory studies to confirm this. Vilas et al. (1994) reported three main-belt asteroids, a Cybele, and a Hilda asteroid with absorption bands in the 0.6-0.65 μm and 0.8-0.9 μm ranges, which could be attributed to saponites (Cloutis et al., 2011a, 2011b). Only one of these objects, 165 Loreley, was observed by the SMASS survey, where it was classified as a Cb and these features not reported (*Bus and Binzel* 2002a, 2002b). None of the objects were observed by the S3OS2 survey (*Lazzaro et al.* 2004), and so confirmation awaits future studies.

The presence of a band near 0.7 μm provides a partial link between the Ch asteroids and CM meteorites: they are the only asteroids and only meteorites that have identified absorption bands at those wavelengths. *Carvano et al.* (2003) performed least-squares comparisons of the asteroidal spectra in the S3OS2 to laboratory spectra of meteorites and concluded that most asteroids with 0.7-μm bands were "very likely" to be related to the CM2 meteorites. The connection between the Ch asteroids and the Pallas-type 3-μm bands is discussed in section 3.2.

*Fornasier et al.* (2014) analyzed the visible-near IR spectra of 600 low albedo asteroids (C, G, B, F, P in the Tholen taxonomy), including 80 that were newly obtained. They found roughly 50% of the 454 C-class asteroids in their sample have the 0.7-μm absorption, compared to only 10% of the 92 B-class asteroids they studied. The F, G, and P asteroids have far fewer objects in their samples (13, 18, and 23, respectively) but all have strong tendencies to either lack the band (~8% of F asteroids and 4% of P asteroids) or have it (100% of G asteroids). They did not separate the P asteroids from the C-complex asteroids for most of their analysis, but the Ps are a small enough fraction of the overall sample they should not skew the results. *Fornasier et al*. found the fraction of hydrated asteroids decreased with decreasing size, though the decrease was not monotonic. They found no correlation between fraction of hydrated asteroids and geometric albedo. However, they concluded that the 0.7-μm band centers for the meteorites are generally at longer wavelengths than for the asteroids, although no conclusion was reached as to why that is the case. They suggested that spectral differences may be attributed to different mineral abundances between CM chondrites and the asteroids, to possible effects due to grain size, or because the meteorite collection may not be fully representative of the aqueously altered asteroid population.

*Rivkin* (2012) used the SDSS to measure the fraction of objects with a 0.7-μm band, first filtering the large set of objects in the Third Release of the Moving Object Catalog to include only those objects with C-like color indices, with a resulting sample size of 3724 observations of 3124 objects. He used two independent techniques, performing a best fit to the average of each C complex component class convolved to SDSS resolution, as well as fitting the distribution of 0.7-μm band depths to two Gaussians (one with zero band depth representing the C asteroids, one with a non-zero band depth representing the Ch asteroids). He estimated 30% of all C-complex asteroids have a band at 0.7 μm, with some evidence of a higher fraction in the middle asteroid belt (2.50-2.82 AU) compared to the inner or outer belt. When including the SMASS and S3OS2 datasets, it appears the Ch fraction of the C complex reaches an overall minimum near H ~12-14. *Fornasier et al.* noted some of the differing conclusions between their work and *Rivkin's*, suggesting that some of the differences may be due to the non-optimal filter placement in the SDSS observations compared to the placement and width of the 0.7-μm band.

All of the 3-μm band shapes discussed above are found in the C complex population. *Rivkin et al.* (2013b) reports that all of their sample of over 30 Ch class asteroids have 3-μm bands consistent with being all "sharp" (or "Pallas") types, with the few possible exceptions attributable to low quality data or misclassified asteroids. However, Pallas types are also present among other C complex classes, as well as non-C classes.

The Japanese AKARI satellite conducted the first spaceborne asteroid spectroscopic survey in the 3-μm region, observing 33 low-albedo asteroids (*Okamura et al.,* 2014a). *Okamura et al.* (2014b) performed Principal Component Analysis (PCA) on these data as well as meteorite and mineral data. While some good matches were found, in general close matches between the AKARI observations and meteorite spectra were not found. However, the depth of the 2.7-μm absorption was found to correlate with the depth of the 0.7-μm absorption, interpreted as meaning that the 2.7-μm absorption is controlled by serpentine abundance.

Work at longer wavelengths has been more limited. However, *McAdam et al.* (2013) analyzed Spitzer Space Telescope data for the 10-12-μm feature discussed above, finding it in seven C-complex asteroids but absent from three D-class Trojan asteroids. Licandro et al. (2012) reported Spitzer spectra covering the 5-14 μm region for 8 Themis family asteroids (including members of several classes in the C and X complexes: *Bus and Binzel* 2002b, *Lazarro et al.* 2004), finding broad emission features of ~2-4%. However, these were not interpreted in terms of phyllosilicate content. Spectroscopic observations of the OSIRIS-REx target asteroid (101955) Bennu from 5-38 μm by Emery et al. (2014) showed no spectral features outside of observational uncertainties.

*3.3.2   B-class asteroids:*
*Ali-Lagoa et al.* (2013) studied the physical properties of B-class asteroids, derived by fitting a thermal model to Wide-field Infrared Survey Explorer (WISE) observations, i.e. effective diameter, beaming parameter, and the albedo ratio $p_{IR}/p_V$, where $p_{IR}$ is the albedo at 3.4 μm as defined in *Mainzer et al.* (2011). *Alí-Lagoa et al.* include in their B-class sample all objects that have a published flat to slightly blue spectral slope in the visible range, i.e. any object that has ever been classified as B-class, including Tholen's F-types, and ambiguous designations, following *Clark et al.* (2010) and *de León et al.* (2012).

By combining the IR and visible albedos with 2.5 µm reflectances from the literature (*Clark et al. 2010*; *de León et al.* 2012) *Ali-Lagoa et al.* obtained the ratio of reflectances at 3.4 and 2.5 µm, from which they find statistically significant indications that the presence of a 3 µm absorption band related to water may be commonplace among the B-types. Using Tholen's B type ECAS classification, *Vilas* (1994) concluded that 30% of those asteroids showed a 0.7 µm absorption, which - using *Howell et al.'s* (2011) correlation - sets a significant lower limit on the presence of water in the B types.

*3.3.3   P-class Asteroids:*
The X-complex asteroids span a range of albedos, as has been known since the 1980s when the Tholen taxonomy recognized the E, M, and P classes were indistinguishable without albedo information and created the X class to hold such objects. The Bus taxonomy (and later Bus-DeMeo taxonomy) does not use albedo information, and has divided the X complex into several classes (Xe, Xk, Xc, X) based on spectral features alone, with imperfect correlations to the Tholen classes. While the Bus classes in the C complex are well-suited to a discussion of hydrated minerals, the Tholen classes provide more convenient discussion of hydrated minerals in the X complex. Therefore, at the risk of some confusion, and understanding that several asteroids are classified in either the X or C complex depending on the specific spectrum analyzed, we use the E/M/P classes defined by Tholen in this section rather than the Bus classes.

The low-albedo X/P-class asteroids fit neatly into the discussion of the C-class asteroids above, and as noted the work of *Fornasier et al.* (2014), following work of others, concluded they belonged to the same alteration sequence as the C-complex asteroids. The largest P-class asteroid, 65 Cybele, has been found to have a Themis-like 3-µm band, interpreted as due to surface ice frost (*Licandro et al.,* 2011a). The asteroid set analyzed by *Takir et al.* (2012) included five P-class asteroids, including two in the Cybele region and two in the Hilda region. These four objects all have rounded 3-µm bands, while the final P asteroid (140 Siwa) did not show any absorption above the noise. The three Cybele/Hilda group C-class asteroids studied by *Takir et al*. included both rounded and sharp members, providing further evidence that spectral slope and 3-µm band shape are correlated, and perhaps connected.

*3.3.4   M-class asteroids:*
Our understanding of M-class asteroids continues to evolve. While the presence of a 3-µm band on several M asteroids (which led to their proposed reclassification as "W-class asteroids" by *Rivkin et al.*) was established late last century (*Jones et al.,* 1990; *Rivkin et al.,* 2000), no consensus on interpretations was reached. The discovery of objects with both a 3-µm band and high radar albedo (*Shepard et al.* 2015) has led to the consideration of impactor contamination as a cause of the 3-µm band on these objects, as noted in Section 1.4, but the band depths on some objects seem too deep to be caused by this process or are on objects with low radar albedo. While work is ongoing in general, we note that some objects have been reclassified into or out of the M asteroid class based on additional spectra or new albedos (for instance, 785 Zwetana was classified as an M by *Zellner et al.,* 1985 but as a Cb by *Bus and Binzel,* 2002a) and that the Lutetia-type 3-µm band shape has only been seen in E- or M-class objects, suggesting that it is not simply caused by contamination by common carbonaceous material in all cases.

Several studies focused on the M-class asteroid 21 Lutetia, a Rosetta flyby target. *Rivkin et al.* (2011) compiled 3-µm observations over several years to conclude Lutetia had a deeper 3-µm band (3-5%) in its southern hemisphere than its northern hemisphere (<2%), and that goethite appeared to be a plausible surface constituent.  The variation observed is consistent with the hemispheric-level variation seen at shorter wavelengths (*Nedelcu et al.* 2007), where Lutetia's spectral slope varies from X-like to C-like.  Several papers by *Birlan et al.* taken together also reported spectral variation on Lutetia consistent with what was seen by *Rivkin et al.*, with *Birlan et al.* (2010) reporting an upper limit of 0.5% for any 3-µm band depth from observations in 2010 compared to a band depth of several percent in 2003-2004 (*Birlan et al.,* 2006).  The Rosetta flyby passed the northern hemisphere of Lutetia in July 2010, and found no absorption in the 3-µm region, though instrumental artifacts make it difficult to judge the actual upper limit value.  *Barucci et al.* (2012), considering all published data, concluded that Lutetia had a chondritic surface, possibly a mixture of enstatite and carbonaceous chondrite. The northern hemisphere of Lutetia stands as a cautionary tale about the difficulties of interpreting featureless spectra. Interestingly, the Alice instrument on Rosetta found a significantly lower UV albedo near 180 nm than 550 nm, and strong reflectance drop from 180 to 160 nm (*Stern et al.,* 2011). The possible interpretations include water ice frost, though there are other possible interpretations, and the survival of frost on Lutetia's surface would be unexpected.

3.4  **Near-Earth Asteroids**

While the Near-Earth Asteroids (NEA) are derived from the main belt, we are aware that the populations are not exact mirrors of one another. The most important difference relevant to this chapter is the relative lack of Ch asteroids in the NEA population relative to their numbers in the main asteroid belt. As we have seen, estimates of the fraction of main belt C-complex asteroids with 0.7-µm bands (or in other words the "Ch fraction") range from ~30-45% (*Rivkin,* 2012; *Fornasier et al.,* 2014). For comparison, *Binzel et al.* (2004) only found one Ch asteroid in their sample of 24 C-complex NEAs. Updated data from the MIT-UH-IRTF "Joint Campaign" finds 6 Ch or Cgh asteroids in a sample of 88 NEOs for which they have data covering the 0.7-µm region ((144901) 2004 WG1, (285263) 1998 QE2, (365246) 2009 NE, 2002 DH2, 2007 YB2, 2012 EG5: *Binzel and DeMeo*, personal communication), a similar (if slightly larger) fraction as found in the 2004 work.  This does not seem to be a size effect—the Mars-crossers in the *Binzel et al.* (2004) sample have a Ch fraction similar to the main-belt value (though the larger Joint Campaign dataset finds a much smaller Ch fraction), and *Rivkin* (2012) found a Ch fraction of 25-40% for the smallest size bin (H of 16-18) in the SDSS data, of similar size to the NEAs.  Perhaps surprisingly, the small fraction of Ch-class NEAs is also inconsistent with the fraction of hydrated meteorites—while it is not obvious how best to compare numbers of meteorites in a consistent way, the fraction of carbonaceous chondrite meteorite falls of the CM group (in which the 0.7-µm band is prevalent: see Section 3) is roughly 33%, again much higher than the 5% seen in the NEO population. A possible explanation for this can be found in the work of *Marchi et al.* (2009), who noted that the chaotic evolution of NEA orbits can lead to low-perihelion eras where surface temperatures exceed the stability temperatures of phyllosilicates. Given the factor of 4-8 difference between the Ch fraction of NEAs vs. main belt asteroids and meteorites, Rivkin (2012) estimated ~75-88% of NEOs have experienced such eras, though this is perhaps a simplistic approach.

Observations of NEAs in the 3-µm region have been rare, with the vast majority of objects too faint to observe. However, absorptions have been found on two NEOs, 1992 UY4 (*Volquardsen et al.,* 2007) and 1996 FG3 (*Rivkin et al.,* 2013a). The band shapes for these objects are not as well-measured as those of brighter asteroids, but 1996 FG3 appears to have a Pallas-type/sharp band. *Volquardsen et al*. interpreted the band shape of 1992 UY4 as similar to that of 375 Ursula, which is a Ceres- or Themis-type, but the observational uncertainties do not rule out a Pallas-type spectrum at the 1-sigma level.

Absorptions in the 3-µm region have also been detected on the S-class asteroids 433 Eros and 1036 Ganymed (*Rivkin et al.,* 2013c), with *Wigton et al*. (2014) also detecting a 3-µm band on 3122 Florence. At this writing, it is not yet clear whether these bands are caused by native material like the phyllosilicates found in the LL chondrite Semarkona, whether they are caused by solar wind interactions with regolith like what is discussed in section 1.5 above, or whether they are due to some other process.

## 3.5 Main Belt Comets (aka Active Asteroids) and Ceres

The discovery of "Main Belt Comets" (MBCs, *Hsieh and Jewitt,* 2006; *Jewitt et al.* this volume) also known as "Active Asteroids" suggests that ice persists near the surfaces of objects in the asteroid belt. Active asteroids are members of the near-Earth or main belt asteroid populations that present comet-like tails, but have orbits that are unlikely to originate from the known comet reservoirs, e.g., the trans-Neptunian belt and the Oort cloud, and are thought to have formed in the main belt instead. These objects eject dust, producing transient comet-like comae and tails. Several dust ejection mechanisms have been proposed: sublimation of ice, rotational fission, ejection by collisions, electrostatic repulsion, thermal fracture, dehydration stresses, and radiation pressure sweeping (e.g. *Bertini* 2011, *Jewitt* 2012, and *Jewitt et al.* this volume), and at least some of the objects are best explained as experiencing processes unrelated to volatiles (e.g. *Jewitt et al.* 2015). Currently, 15 active asteroids are known, two of these are NEAs (3200 Phaethon and 4015 Wilson-Harrington) while the other 13 are in the main belt.

The detection of repetitive dust ejection events in some MBCs (e.g. 133P/Elst-Pizarro and 238P/Read, see *Hsieh et al.* 2004, *Jewitt et al.* this volume and references therein) and the indication that in these and other MBCs (e.g. 238P/(300163) 2006 VW139 , P/2010 R2 (La Sagra) and P/2010 T1 (PANSTARRS) , see *Jewitt et al.* this volume) dust ejection lasted for several months strongly suggest that ice sublimation is their preferred ejection mechanism (*Hsieh et al.* 2004). Interestingly, three MBCs, 133P, 176P, and 288P, are members of the Themis dynamical family and share spectral properties with other asteroids in this family (see *Licandro et al.,* 2011b, 2013, and *Hsieh et al.* 2012), indicating they have a common origin. Furthermore, water ice was discovered on the surface of the largest member of this family, (24) Themis (*Campins et al.,* 2010, *Rivkin and Emery,* 2010). Because the water ice on the surface of Themis is short-lived, it likely is resupplied from its interior, and this also supports the hypothesis that there is a water ice reservoir below the surface of some MBCs. Thermal models are consistent with the survival of very shallow subsurface water ice over the age of the Solar System in main belt asteroids (*Schorghofer,* 2008).

One interesting question regarding Themis family MBCs is: why are only some Themis family asteroids active? If activity in MBCs is indeed driven by ice sublimation, this ice may normally be protected from sunlight by a poorly conducting surface layer and activity could start with a collision penetrating this layer, exposing the ice to sunlight (*Hsieh and Jewitt*, 2006; *Schorghofer,* 2008, *Capria et al.* 2012). Recent dynamical results are consistent with this proposed mechanism; more specifically, the discovery of a very young sub-family within the Themis family, called the Beagle family, and the fact that MBC 133P is also a member of the Beagle sub-family (*Nesvorný et al.,* 2008). The relatively recent formation event of the Beagle family may have produced fragments, such as 133P, with shallow ice that can be exposed by small (and thus more frequent) collisions. However, since the other MBC in the Themis family, 176P, is not a member of the Beagle family, multiple collisional events or perhaps surface disruption following YORP spin-up are needed to explain their activity (*Campins et al.,* 2012). Another factor that may help explain why only a few Themis family members show activity is a heterogeneous set of compositions among small Themis fragments. This issue can be addressed by focused spectroscopic studies of this family (Section 4.7)

It is also suggestive that all the MBCs with known spectra are primitive asteroids. More specifically, 133P, 176P, 288P, (3200) Phaethon and (4015) Wilson-Harrington all belong to the C-complex in the Tholen taxonomy (*Licandro et al.,* 2007, 2011b, and 2013), and asteroid 596 Scheila (whose activity is attributed to an impulsive event like a collision, e.g. *Moreno et al.,* 2011) is a D-type asteroid (*Licandro et al.,* 2011c).

A particularly interesting and related result is the detection by the Herschel Space Telescope of water vapor emission around asteroid (1) Ceres (*Küppers et al.,* 2014). Models of Ceres' interior include a near-surface ice mantle above a rocky core (*Thomas et al.,* 2005, *McCord and Sotin*, 2005) with the ice separated from the vacuum of space by either a lag deposit similar to comets or a primordial crust. The water emission seen by Herschel is variable, suggesting a connection with surface features, and correlated with solar distance. While Ceres is not considered an MBC (no dust emission or tail has ever been seen), it is possible that MBCs and Themis originated in Ceres-like bodies that were catastrophically disrupted (*Castillo-Rogez and Schmidt*, 2010). Such a scenario could increase the ice fraction among some fragments and also maintain it near the surface in small objects like the MBCs, while still leaving it accessible for activation and sublimation via impacts.

## 3.6 Related populations

The Trojan asteroids and satellites of Mars are discussed in detail in the chapters by *Emery et al.* and *Murchie et al.* in this volume. However, it is worth briefly discussing them again here, along with the irregular satellites of Jupiter, in the context of their hydrated or icy mineralogy. The Cybele and Hilda asteroids are considered in the context of the main belt asteroids throughout this chapter, though dynamical studies suggest their origins may lie outside the main belt and with the other objects discussed below.

The Trojan asteroids of Jupiter are dominated by members of the low-albedo P and D classes in the Tholen taxonomy: *DeMeo and Carry* (2013) estimate the Trojan population is 2/3 D-class asteroids by mass, with P-class asteroids making up the vast majority of the remainder. In a follow-up, *DeMeo and Carry* (2014) looked at the mass distribution with size, finding the D:P:C ratio varies wildly in the Trojan population, with Ds dominant in the 50-100 km size range but the D:P fraction more equal outside that range. Focused studies of Trojan asteroids by *Dotto et al.* (2006), *Fornasier et al.* (2007), and *De Luise et al.* (2010) found no objects with the 0.7-µm band seen in CM meteorites and Ch asteroids (Section 2.1), even among objects with spectral slopes that place them in the C complex. *Emery and Brown* (2003) studied 20 Trojan asteroids in the 3-µm region, also finding them featureless. This is consistent with a lack of hydrated minerals, but the available spectra are necessarily of lower quality than main-belt objects of similar size. *Emery and Brown* (2004) find that the absence of absorptions in the 3-µm region limits the abundance of $H_2O$ ice to a few wt% (at the surfaces), but up to 10–20 wt % hydrated minerals could be present and not seen in those data. *Cruikshank et al.* (2001) noted that 624 Hektor, the largest Trojan, could have 3 wt% ice or 40% hydrated minerals on its surface given observational uncertainties. If hydrated minerals are present on Trojan asteroids, their detection awaits observations from spacecraft visits or with larger or more capable telescopes than have been available to this point raft.

*Vilas et al.* (2006) compiled *BVRI* , ECAS filter photometry, and narrowband photometry, and CCD reflectance spectra of Jovian irregular satellites, determining whether a 0.7-µm band was present. Of their sample of 16 targets, 11 tested positive for a 0.7-µm band. However, each of the groupings represented (Himalia prograde, Pasiphae retrograde, Carme retrograde) had both members that tested positive and that tested negative. The sample also contained both C types that tested positive and negative and P/D types that tested positive and negative. New CCD reflectance spectra (*Vilas* 2010, 2013) confirm that the 0.7-µm feature is present in all Himalia group objects in their sample as well as some objects from the Pasiphae group.

Himalia, the largest of the Jovian irregular satellites, has been observed from the ground and space. *Jarvis et al.* (2000) interpreted spectral and spectrophotometric data for Himalia to include a 0.7-µm band, though observational uncertainties were large. *Chamberlain and Brown* (2004) analyzed Cassini Visual and Infrared Mapping Spectrometer (VIMS) observations of Himalia, finding the "suggestion" of a 3-µm band, though large uncertainties frustrated a detailed spectral fit. More recently, *Brown and Rhoden* (2014) detected a 3-µm absorption on Himalia using the Keck telescope, finding a spectral shape that is a good match to the Themis/rounded type found in main-belt asteroids. While the Vilas et al. and Jarvis et al. findings of a 0.7-µm band would lead one to expect a 3-µm band on Himalia, a Pallas-type/sharp band shape would be expected (see Section 3.2). The finding of a Themis-type band shape led *Brown and Rhoden* to suggest additional observations of Himalia were needed to confirm the presence of a 0.7-

µm band.  An additional possible explanation for the mismatch between the 3-µm band shape seen on Himalia based on expectations from main-belt asteroids may be found in the work of *Clark et al.* (2012).  They suggest that a band at 0.67 µm seen on Iapetus by *Vilas et al.* (1996) may be due to nanophase hematite and/or iron. This may be a way of explaining a co-occurrence of the 0.7-µm and non-Pallas-type 3-µm bands, though obviously more work is needed.

Finally, *Fraeman et al.* (2014) used data from the Mars Reconnaissance Orbiter's (MRO) Compact Reconnaissance Imaging Spectrometer for Mars (CRISM) to detect absorptions centered near 0.65 and 2.8 µm on Phobos and Deimos, the latter interpreted as due to hydrated minerals and with a band shape similar to the Pallas/sharp types.  The band at 0.65 µm was interpreted as either due to hydrated minerals (and the equivalent of the 0.7-µm band discussed above) or, again, as due to nanophase materials.  It is worth noting that while all of Deimos' surface and most of Phobos' surface had these absorptions, the areas of Phobos associated with Stickney and its ejecta have a more shallow spectral slope than the rest of Phobos and lack these absorptions, and are apparently anhydrous. Previous Earth- and space-based observations of Phobos and Deimos did not find convincing evidence for hydrated minerals (*Murchie and Erard*, 1996; *Rivkin et al.,* 2002a; *Gendrin et al.,* 2005.)

The Trojan asteroids, irregular satellites of the giant planets, and perhaps even Phobos and Deimos are thought to derive from the transneptunian object (TNO) population in Nice Model interpretations. TNOs are well-known to have icy surfaces and have a variety of spectral slopes similar to primitive asteroids. Other than Pluto, however, they are too faint to be observed in the 3-µm region to study non-ice hydrated minerals. Aqueous alteration products have been reported in the 0.7-µm region on some TNOs and the related Centaur population (*De Bergh et al.* 2004), though these interpretations have not been confirmed (*Fornasier et al*. 2009) and band centers are different from what is seen on Ch asteroids (perhaps they are due to nanophase materials rather than phyllosilicates, as suggested above for Himalia). *Barucci et al.* (2008) reviews the evidence for TNO aqueous alteration from visible spectra.

3.7  **Rotational Variation**

To investigate the correlation of the 0.7- and 3-µm absorption bands, *Howell et al. (2005)* reported  rotationally resolved spectra of several asteroids to test whether variability with rotation could explain previous inconsistent observations. When simultaneous observations were not possible, full rotational coverage was obtained to explore whether spectral variation was present. The sample was not unbiased, since objects with inconsistent spectral observations were specifically targeted, so the rate of variability seen may not be typical. However, band depth variations ranging from 1-5% were found in most of the objects studied, suggesting that variation is common, if not ubiquitous. Variation in band depth is now thought to explain all of the cases where an asteroid with a 0.7-µm band was not seen to have a corresponding 3-µm band.  For instance, *Lebofsky* (1980) found that the Ch asteroid 554 Peraga did not have a 3-µm absorption, but due to the wavelength he used as the continuum to define the band, and the low spectral resolution of their photometry, those observations did not have the precision required to detect the band that is now seen.

*Rivkin and Volquardsen* (2010) observed Ceres over its entire rotation, finding band depths to be shallower at the longitudes associated with bright regions in the HST maps. This could be due to a smaller concentration of brucite and carbonates, or plausibly consistent with a variation in particle size. The variation seen and the precision of the data led to an estimate that brucite-free areas 200 km in diameter (at the equator) or larger would have been detected, and that if the ~3% variation in band depth were entirely due to the bright regions, those bright regions would have band depths of ~15-17% compared to the ~20-21% global average band depth.

Improved instrumentation has greatly expanded the number of asteroids observable at 3 µm, and the sensitivity of the band depth that is detectable. However, the interpretation of the Vesta low albedo surface material as impactor ejecta (Section 1.6) might lead one to think that contamination may be a common occurrence and explains most if not all observations of variation. The nature and depth of the 3-µm band this is observed suggests that this is not the case. Ground-based observations of Vesta show a much weaker band than the 10-20% band seen on most Ch asteroids. While other bright S-type objects (and members of other taxonomic classes not expected to have abundant hydrated surface materials) may show possible bands of a few percent, none show the deep, obvious bands of the Ch or even many of the W-class objects. More work certainly needs to be done to clarify this issue, but the current evidence suggests that there are asteroids composed of abundant hydrated silicates, formed in place, parents of CM and CI chondrites, among others. There may also be traces of impactor material detectable on asteroids without native hydrated silicates in their interiors, but careful groundbased spectroscopy can often distinguish them from each other. Improving these methods and finding robust tests of hydrated composition will certainly be a high priority in the future and an important goal of future missions.

## 3.8   Family studies

Members of asteroid families are the fragments of larger, collisionally disrupted or cratered parent bodies; spectroscopic studies of family members provide information about the interior of the parent body. A trio of chapters in this volume (by *Nesvorný et al.*, *Masiero et al.,* and *Michel et al.*) addresses various aspects of asteroid family studies. Observations of the Themis family serve to illustrate the benefits of characterizing hydration and other properties of family members.  The Themis family is one of the most numerous and statistically reliable in the asteroid belt. It is approximately 3.2 AU from the Sun in low-inclination orbits and, as discussed in Section 3.5, it contains three MBCs. Spectroscopic studies of this family reveal an interesting picture. *Florczak et al.* (1999) obtained visible spectra (0.49 to 0.92 µm) of 36 members of this family, and found that about 50% of their sample showed evidence of aqueous alteration, indicating that the parent body was sufficiently warm to mobilize water. Their results also suggest that the percentage of asteroids showing aqueous alteration may decrease with the diameter of the objects; however this needs confirmation as it could be due to observational biases including the difficulty of detecting the 0.7-μm band in the fainter targets. In addition, *Campins* (2014) pointed out that Themis family fragments show clear spectral diversity in the near-infrared (1-2.5 μm), with individual objects falling in four of the five groups of B-type near-infrared spectra identified by *de León et al.* (2012). Wide variety in spectral slopes was also seen in the Themis family by *Fornasier et al*. (2014). Moreover, the 3-μm band on (24) Themis itself is indicative of water ice as opposed to hydrated minerals, as noted. The above results support at least some compositional diversity in the parent body of the Themis family and the distribution of compositions could be attributed to fragments coming from different depths in the original parent body. This tantalizing result and other aspects of family structure and composition can be addressed by new spectroscopic surveys, such as ESA's Gaia mission (*Campins et al.,* 2012) and targeted ground and space based studies.

Family studies will also be important for understanding the nature of the different 3-μm types. While families are known to be realtively homogeneous in terms of their Tholen or Bus-DeMeo classifications, it is not yet clear how homogeneous they will be in terms of their hydrated mineralogies. The Hygiea family in particular should provide an interesting test: *Rivkin et al.* (2014a) argued that the lack of a Ceres family is supporting evidence that it has an icy shell covered by a lag deposit so that any collisional debris would quickly sublime away. It is tempting to assign a similar interior structure to Hygiea given its surface similarity to Ceres, but the existence of a Hygiea collisional family is at odds with the Ceres family scenario. Observations to determine whether Hygiea family members share a similar 3-μm band shape or not will be critical in interpreting whether Hygiea is an interloper in its own family.

# 4   Discussion

While much work is still to be done, and some key datasets are not yet in the literature or in some cases even yet obtained, the outlines of an overall scenario can be discerned. The strong correlation between the 0.7-μm band and the Sharp/Pallas-type 3-μm band shape suggests that the process creating the minerals responsible for the former also creates minerals responsible for the latter. However, because some Pallas types can be found without a 0.7-μm band, the situation is likely not as simple as one mineral being responsible for both bands in all cases. Similarly, the presence of Themis-type band shapes in members of the C, P, and B classes suggests that whatever is causing the different spectral slopes of these objects is independent of the hydrated mineral composition, and the apparent concentration of the Ceres-type bands in large asteroids, if it is not related to poorer-quality data for smaller (and on average fainter) objects, is suggestive of similar histories for those objects (although see Section 3.8). Furthermore, the fact that only Pallas-type bands are found in the meteorite collection is also likely an important clue.

The process or processes that changed aggregations of pebbles and dust (with or without ice) into solid rock is not well understood (*Consolmagno et al.* 2003, *Connolly, personal communication*). Indeed, the process by which planetesimals are assembled has been a subject of much recent interest, with recent evidence suggesting that 100-km scale objects are created directly from cm-size building blocks, without intermediate stages (*Morbidelli et al.,* 2009).

One can imagine scenarios that, while at this point provisional, account for all of these factors: carbonaceous/low-albedo 100-km scale objects form as dust/pebble/ice aggregates into objects we would now call Themis types. The objects that have internal heat sufficient to melt ice undergo aqueous alteration, which can include exothermic reactions that lead to a self-perpetuating process. The aqueous alteration process lithifies bodies, and creates the minerals that exhibit Pallas-type bands. Objects that are sufficiently large and/or contain sufficient heat can differentiate with an icy crust/mantle over a rocky core. The alteration processes associated with this pathway may create the brucite and carbonates seen on the Ceres types. The Themis types remain as they were at formation, unlithified and so unable to generate meteorites. The Ceres types may not create meteorites because they have near-surface ice, and large impacts may only create icy families that are short-lived against sublimation instead of the rocky families created by other parent bodies (see *Rivkin et al.,* 2014a). In this scenario, only the Pallas types, lithified by aqueous alteration, can generate low-albedo, carbonaceous material strong enough to survive passage through the Earth's atmosphere to be collected as meteorites.

This scenario has potential inconsistencies—for instance, it's not clear if dynamical families can be created in situations where meteorites are not predicted (for instance, with the Themis family), and it's not obvious that Hygiea and Ceres, both of which have very similar spectra in the 3-μm region, will necessarily have had similar histories and have similar interiors. Nevertheless, the ongoing work in understanding the hydrated mineralogies of the asteroids seem likely lead to paradigms at least broadly similar to what is outlined above.

# 5   Future work

We have spent the previous pages discussing the current state of affairs with regard to water/hydroxyl in the asteroid and meteorite populations. The upcoming decade promises further progress on many fronts, from Earth-based observations using new capabilities and facilities to *in situ* observations by spacecraft and sample returns.

## 5.1   Observing from the stratosphere and above

As noted above, the Earth's atmosphere places limitations on groundbased studies of asteroidal water/OH, making some wavelengths basically unobservable. The capability to observe from above the Earth's atmosphere would revolutionize the work presented here, allowing OH band depths and band centers to be directly measured instead of inferred or modeled. The spaceborne telescopes of the late $20^{th}$/early $21^{st}$ century generally did not cover the 3-μm region, or had broadband filters that only allowed indirect measurements, but the next generation of facilities should provide improved coverage.

*5.1.1 SOFIA:*
The SOFIA project has been fully commissioned, with the FLITECAM instrument providing spectroscopic data with spectral resolution ($\lambda/\Delta\lambda$) of 1300 from wavelengths 1.14-4.07 μm, with some gaps in coverage. Its typical flight altitude of 11-14 km puts it above 99% of the Earth's water vapor, making it well suited to observations in the 3-μm region. However, it is unclear how many asteroids will be observable due to proposal pressure over SOFIA's planned lifetime, and budget pressures threaten to shorten the mission lifetime to a much shorter one than originally expected.

*5.1.2 Balloon-borne Observations and Suborbital Flights*
New efforts to use balloon-borne telescopes to study small bodies began with the effort mounted to observe Comet ISON in 2013 ("BRRISON"). Very high-altitude balloons show the promise of providing a long-term platform to study astronomical phenomena above much of the Earth's atmosphere, thus reducing, if not quite fully eliminating, the effects of the Earth's atmosphere on IR observations (*Young et al.*, 2013), although pointing stability of the balloons remains an ongoing concern. The BRRISON and BOPPS missions included asteroids in their observing plans and Ceres was observed with UV/visible and infrared instruments during the BOPPS flight, which occurred as this was being written.

Concurrently, telescopes are being developed for commercial suborbital spacecraft that will operate above the Earth's atmosphere. These should be capable of making rapid timescale observations, and can sample UV through the NIR spectral region (*Vilas et al.*, 2013) eliminating atmospheric contamination effects, but are more limited in observational duration. These techniques also hold the promise of expanding observations of water-bearing asteroids while minimizing atmospheric contamination.

*5.1.3 James Webb Space Telescope*
The James Webb Space Telescope (JWST) has similar potential to revolutionize asteroidal studies. Its Near-Infrared Spectrograph (NIRSPEC) instrument has spectroscopic capabilities in the 1-5 μm spectral region, and the Mid-Infrared Instrument (MIRI) provides low-resolution spectroscopy ($\lambda/\Delta\lambda \sim 100$) from 5-14 μm and medium-resolution spectroscopy ($\lambda/\Delta\lambda \sim 2200-3500$) and imaging in nine bands from 5-28 μm. The anticipated sensitivity for JWST should allow high quality (S/N >30 in $10^4$ seconds) measurements in the 3-μm region to be made for low-albedo main-belt asteroids of ~3 km diameter, that size rising to ~10 km in the Trojan clouds (or for the irregular satellites of Jupiter). Unfortunately, the current capabilities planned for JWST do not include the ability to track objects moving faster than Mars, which precludes observations of NEOs at arbitrary times (for instance, near perihelion and/or away from quadrature). It is possible, however, that certain NEOs of interest can be observed at specific times when their apparent rates of motion are slower (if such times exist). As with SOFIA, proposal pressure for JWST may be the greatest impediment to asteroidal studies.

## 5.2 Spacecraft visits to asteroids

In addition to spacecraft telescopically observing asteroids, at least three spacecraft will be visiting asteroids and making close-up observations. Dawn, mentioned above in the context of Vesta, will be visiting Ceres near the time this book is published. We expect it to make great progress toward answering several of the questions posed here. The other two spacecraft, Hayabusa-2 and OSIRIS-REx, are intended as sample return missions to primitive asteroids. However, both missions also carry infrared spectrometers that will allow observations in the wavelength regions discussed above to provide context for the samples.

*5.2.1 Dawn at Ceres*
The Dawn mission team has contrasted the water-rich nature of Ceres to the water-poor nature of Vesta, and several of the mission goals at Ceres relate to water, including mapping hydrogen abundance using the neutron spectrometer. More pertinent to this chapter, the VIR visible and near-IR spectrometer covers the 0.25-5 µm spectral range, including both the 0.7-µm and 3-µm region. The mission plan includes obtaining half of the VIR frames with a spatial resolution of 400 m or better, with the remainder at a spatial resolution of 1.6 km or better. While Earth-based observations have not detected ice on Ceres' surface, the higher spatial resolution available in the Dawn data will be able to detect much smaller amounts of surface ice (stable at very high latitudes) or place much stronger upper limits on ice abundance than are currently available. It will also provide context for the Herschel observations of water sublimation from Ceres (*Küppers et al.* 2014), perhaps by identifying surface features related to sublimation.

*5.2.2 Hayabusa-2*
The Hayabusa-2 mission is due to be launched in 2015 by the Japanese space agency JAXA, targeting the low-albedo asteroid 1999 JU3 for a sample return mission. Observations of 1999 JU3 place it in the C class (Pinilla-Alonso et al. 2013). Besides the returned sample, Hayabusa-2 carries two instruments that will be able to detect and characterize hydrated minerals on the asteroidal surface: a multiband imager (ONC-T), which would allow measurement of the 0.7-µm band if present at spatial resolution of 2 m/pixel, and a near-IR spectrometer (NIRS3), which covers a wavelength range of 1.8-3.2 µm. The goal of NIRS3 is to allow hydrated mineral quantities to be estimated to ~1-2% (Iwata et al. 2013), and the spatial sampling is ~40m/spectrum at the 20 km "home position" altitude, improving to 2 m/spectrum during one low-altitude mission phase.

*5.2.3 OSIRIS-REx*
As noted above, NASA's OSIRIS-REx mission will return samples to Earth from the NEA 101955 Bennu (*Lauretta et al., in press*). The OSIRIS-REx spacecraft will be well-instrumented to carry out a year-long campaign to characterize Bennu and select a sampling site. The visible and infrared spectrometer (OVIRS) will map the spectrum of the asteroid from 0.4 to 4.2 µm. These data will be well-suited for direct searches for the 0.7- and 3-µm hydration features. The thermal emission spectrometer (OTES) will similarly map Bennu from 6 to 50 µm, providing complementary spectral information in the mid-infrared for mineralogical characterization. Thermal mapping at different times of day and points in the orbit around the sun will enable monitoring of temperatures and derivation of spatially-resolved thermal inertias, which can both be used to understand the stability of water (in its various forms) on and below Bennu's surface as well as measure mid-IR bands associated with phyllosilicates (Section 2.3). The imagers will be used for sensitive searches for extended emission and/or plumes, which, if present, could suggest either thermal fatigue- or volatile-related release of dust. Most importantly, the samples returned to Earth will be analyzed in detail, providing a window into the make-up of primitive solar system material and excellent ground-truth for the interpretation of remote observations.

## 5.3  Large ground-based telescopes

The prospects for larger ground-based telescopes provide additional opportunities for observing smaller and more distant asteroids in the future. Planned telescopes in the 20-meter and larger class like the Giant Magellan Telescope (GMT: 25-m), Thirty Meter Telescope (TMT: 30-m) and European Extremely Large Telescope (E-ELT: 39-m) will begin operation late this decade and early next decade. Instrumentation on the TMT and GMT, expected to be the first ones completed, will not cover the 3-µm region at first and will only cover the 0.7-µm region at very high spectral resolution (http://tmt.org/sites/default/files/TMT-Instrumentation-and-Performance-Handbook.pdf, http://www.eso.org/sci/meetings/2013/EELT2013/talks/Jacoby.pdf ). Instrumentation at the E-ELT is yet to be finalized. Despite this, we expect members of the community to make use of these facilities in ways that are not necessarily obvious at the moment.

# 6   Open Questions

1. *Can the different 3-µm types be found on a single body?*   Did the aqueous alteration process create spectrally distinct minerals at different times? Were some minerals more commonly created with depth? Is the 3-µm type a function of starting material?
2. *To what extent is impactor contamination important on small bodies?* If contamination by carbonaceous impactors has caused the 3-µm band on Vesta, is it also potentially important on large (or small) S-class asteroids? What does the existence of different 3-µm types and the correlation of the 0.7-µm band and the Pallas-type 3-µm band tell us about how widespread impactor contamination may be? Can we identify this process on low-albedo objects as well, and should contamination by ordinary chondrite impactors also be expected?
3. *How do the outer solar system small body populations (Trojans, irregular satellites, comets) fit into the overall picture?* At the time of *Asteroids III*, it was thought that asteroids and comets were disjunct sets and the Hilda and Trojan asteroids were in effect an outward extension of the asteroid belt. At this writing, there appear to be ample objects with physical natures that are partially cometary and partially asteroidal. Do the Trojans and irregular satellites represent material that is distinct from both asteroids and comets, or are they simply typical comets that experienced transport early in solar system history?

4. *Does the solar wind create appreciable amounts of OH on small body surfaces?* The water brought to Earth by asteroidal impacts may not only have come on carbonaceous bodies—if significant amounts of OH can be created by the solar wind, most asteroidal impactors regardless of composition will have brought in some water, if only a trace. In addition, while the timescale for mining water from asteroids commercially is likely decades away, the techniques needed to identify good targets will be critically dependent upon knowing whether any OH detected is on the very surface or likely to be representative of the interior.
5. *What is responsible for the varying appearance of the 0.7-μm band in asteroids/meteorites that otherwise appear similar at 3 μm?* The frequent presence of 0.7-μm bands in CM meteorite spectra leads to a reasonable linkage between that meteorite group and the Ch asteroids, as discussed several times above. However, there is a significant fraction of C-complex asteroids without the 0.7-μm band but with a Pallas-type 3-μm band. What is their association with the Ch asteroids?
6. *What variation in band shape is present in each of the 3-μm types, and what is causing that variability?* Are the differences in band position in the Themis types, or the presence/absence of shoulders in Pallas-type bands indicative of important compositional differences?
7. *How can we best account for volatiles hidden from the surface?* The MBCs offer examples of the importance of indirect measurements of water ice when targets are faint. Similarly, the lack of any water or hydroxyl absorption in Rosetta measurements of Comet 67P (*Capaccioni et al.,* 2015) indicates that spectral evidence for volatiles may be absent even if high spatial resolution and high-quality data are available. It is possible that many (most?) undifferentiated low-albedo objects in the outer asteroid and beyond have much more ice than their surface spectra betray, and depending on surface conditions any ice table may be deeper than even neutron spectroscopy can reach.

# 7   Conclusions

Since the publication of *Asteroids III*, the water content of small bodies has been recognized as of great importance to their study, with robotic and human exploration of water-rich asteroids assigned high priority by the international planetary science community. New instruments and capabilities in laboratories and at telescopes have allowed better data to be obtained and the interfering effects of the Earth's water to be minimized. The discovery of the "activated asteroids" and of ice at asteroidal surfaces has demonstrated that free water is still present today in significant amounts. Work by many authors has linked the Ch asteroid class to the CM meteorite group sufficiently firmly that the type of statistical and mineralogical analysis that has been commonplace for ordinary chondrite and basaltic achondrite analogs can now begin for the carbonaceous chondrites. Variety in the hydrated mineralogy of asteroids has been seen, and we are starting to associate that variety with particular asteroid sizes and locations, with an apparent path toward divining their histories. In this effort the ground-based remote sensing that we have historically used to study asteroids will be augmented in the near future with observations by airborne and spaceborne platforms, visits to hydrated asteroids, and ground truth via sample return. By the time *Asteroids V* is released, presumably sometime in the 2020s or early 2030s, it seems likely that the state of knowledge of the hydrated minerals on asteroids will not be containable in a single chapter.

*Acknowledgments:* ASR acknowledges the long-standing support of the NSF and NASA Planetary Astronomy programs and the NASA NEO Observations program. HC acknowledges support through the NASA NEOO and Solar System Observations programs and NASA's OSIRIS-REx mission through a contract with the University of Arizona. JPE and DT acknowledge NASA Planetary Astronomy grant NNX08AV93G, and DT additionally acknowledges support from NASA Cosmochemistry grant NNX10AH48G (Harry Y. McSween Jr, PI). We all thank NASA IRTF staff for their assistance with asteroid observations, and note the IRTF is operated by the University of Hawaii under Cooperative Agreement No. NCC 5-538 with the National Aeronautics and Space Administration, Office of Space Science, Planetary Astronomy Program. Thanks to the people of Hawai'i for allowing observers to visit and build upon the sacred mountain of Mauna Kea for the purposes of advancing science.

# References

Alí-Lagoa, V.; de León, J.; Licandro, J.; Delbó, M.; Campins, H.; Pinilla-Alonso, N.; Kelley, M. S; (2013). Physical properties of B-type asteroids from WISE data, A&A 554; .A71, 16

Arnold, S. M.; (1979). Rapid photometry of cataclysmic variable stars, Ph.D. Thesis Rochester Univ., NY,

Barucci, M. A., Brown, M. E., Emery, J. P., and Merlin, F. (2008). Composition and surface properties of transneptunian objects and centaurs. The Solar System Beyond Neptune, 143.

Barucci, M. A.; Belskaya, I. N.; Fornasier, S.; Fulchignoni, M.; Clark, B. E.; Coradini, A.; Capaccioni, F.; Dotto, E.; Birlan, M.; Leyrat, C.; (2012). Overview of Lutetia's surface composition, Planetary and Space Science, 66; 23-30

Beck, P., Quirico, E., Montes-Hernandez, G., Bonal, L., Bollard, J., Orthous-Daunay, F.-R., Howard, K. T., Schmitt, B., Brissaud, O., and F. (2010). Hydrous mineralogy of CM and CI chondrites from infrared spectroscopy and their relationship with low albedo asteroids. Geochim. Cosmochim. Acta, 74:4881–4892.

Beck, P., Quirico, E., Sevestre, D., Montes-Hernandez, G., Pommerol, A., and Schmitt, B. (2011). Goethite as an alternative origin of the 3.1 μm band on dark asteroids. A&A, 526(A85):A85.

Beck, P.; Garenne, A.; Quirico, E.; Bonal, L.; Montes-Hernandez, G.; Moynier, F.; Schmitt, B; (2014). Transmission infrared spectra (2-25 μm) of carbonaceous chondrites (CI, CM, CV-CK, CR, C2 ungrouped): Mineralogy, water, and asteroidal processes, Icarus, 229; 263-277

Bertini I. (2011). Main Belt Comets: A new class of small bodies in the solar system. Planetary and Space Science, 59: 365-377

Bertrand, M., van der Gaast, S., Vilas, F., Hörz, F., Haynes, G., Chabin, A., Brack, A., and Westall, F., (2009).The Fate of Amino Acids During Simulated Meteoritic Impact, *Astrobiology 9*, 943 (2009).

Binzel, R. P., Rivkin, A. S., Stuart, J. S., Harris, A. W., Bus, S. J., and Burbine, T. H. (2004). Observed spectral properties of near-Earth objects: results for population distribution, source regions, and space weathering processes. Icarus, 170:259–294.

Birlan, M., Vernazza, P., Fulchignoni, M., Barucci, M. A., Descamps, P., Binzel, R. P., and Bus, S. J. (2006). Near infra-red spectroscopy of the asteroid 21 Lutetia. I. New results of long-term campaign. A&A, 454:677–681.

Birlan, M., Nedelcu, A., Vernazza, P., Binzel, R., Carry, B., DeMeo, F., Barucci, A., and Fulchignoni, M. (2010). 21 Lutetia: Groundbased Near-infrared Spectroscopy Prior The Rosetta Flyby. In AAS/Division for Planetary Sciences Meeting Abstracts #42, volume 42 of Bulletin of the American Astronomical Society, page 1050.

Bishop, J. L. and Pieters, C. M. (1995). Low-temperature and low atmospheric pressure infrared reflectance spectroscopy of mars soil analog materials. 100:5369–5379.

Bottke, W. F., Nesvorný, D., Grimm, R. E., Morbidelli, A., and O'Brien, D. P. (2006). Iron meteorites as remnants of planetesimals formed in the terrestrial planet region. Nature, 439:821–824.

Bottke, W. F., Nesvorny`, D., Vokrouhlicky`, D., and Morbidelli, A. (2010). Collisional evolution of the irregular satellites. The Astronomical Journal, 139:994–1014.

Bottke, W.F., Vokrouhlicky, D., Minton, D., Nesvorný, D., Morbidelli, A., Brasser, R., Simonson, B., and Levison, H.F. (2012) "An Archean heavy bombardment from a destabilized extension of the asteroid belt" Nature v. 485, 78-81

Bottke, W. F., Vokrouhlicky`, D., Nesvorny`, D., and Moore, J. M. (2013). Black rain: The burial of the galilean satellites in irregular satellite debris. Icarus, 223(2):775–795.

Britt, D., Schelling, P., Consolmagno, G., and Bradley, T. (2014). Space weathering on volatile rich asteroids. In Lunar and Planetary Institute Science Conference Abstracts, volume 45, page 2067.

Brown, M. E. and Rhoden, A. R. (2014). The 3 μm Spectrum of Jupiter's Irregular Satellite Himalia. Astroph. J. Lett. 793:L44.

Browning, L. B., McSween, H. Y., and Zolensky, M. E. (1996). Correlated alteration effects in CM carbonaceous chondrites. Geochem, Cosmochem. Acta. 60:2621–2633.

Burbine, T. H., McCoy, T. J., Meibom, A., Gladman, B., and Keil, K. (2002). Meteoritic parent bodies: Their number and identification. In Bottke, W., Cellino, A., Paolicchi, P., and Binzel, R. P., editors, Asteroids III, pages 653–667. University of Arizona Press, Tucson.

Bus, S. J. and Binzel, R. P. (2002a). Phase II of the Small Main-Belt Asteroid Spectroscopic Survey: A Feature-Based Taxonomy. Icarus, 158:146–177.

Bus, S. J. and Binzel, R. P. (2002b). Phase II of the Small Main-Belt Asteroid Spectroscopic Survey: The Observations. Icarus, 158:106–145.

Campins, H., Hargrove, K., Pinilla-Alonso, N., Howell, E. S., Kelley, M. S., Licandro, J., Mothé-Diniz, T., Fernandez, Y., and Ziffer, J. (2010). Water ice and organics on the surface of the asteroid 24 Themis. Nature, 464:1320–1321.

Campins, H.; de León, J.; Licandro, J.; Kelley, M. S.; Fernández, Y.; Ziffer, J.; & Nesvorný, D. (2012) Spectra of asteroid families in support of Gaia. Planetary & Space Sci. 73, pp. 95-97.

Campins, H. (2014) Volatiles in small bodies. In Proceedings of Asteroids, Comets, Meteors 2014, Muinonen et al., Eds..

Capria. M.T., Marchi, S., de Sanctis, M.C., Coradini, A., Ammannito, E. (2012), The activity of main belt comets. A&A, 537A, 82C.

Carvano, J. M.; Mothé-Diniz, T.; Lazzaro, D. (2003). Search for relations among a sample of 460 asteroids with featureless spectra, Icarus, 161; 356-382.

Castillo-Rogez, J. C. and Schmidt, B. E. (2010). Geophysical evolution of the Themis family parent body. Geophys. Research Lett., 37:10202.

Chamberlain, M. A. and Brown, R. H. (2004). Near-infrared spectroscopy of Himalia. Icarus, 172:163–169.

Chapman, C.R., Morrison, D., Zellner, B.H., (1975). Surface properties of asteroids – A synthesis of polarimetry, radiometry, and spectrophotometry. Icarus 25, 104–130.

Clark, B. E., Ziffer, J., Nesvorny, D., Campins, H., Rivkin, A. S., Hiroi, T., Barucci, M. A., Fulchignoni, M., Binzel, R. P., Fornasier, S., DeMeo, F., Ockert-Bell, M. E., Licandro, J., and Mothé-Diniz, T. (2010). Spectroscopy of B-type asteroids:

Subgroups and meteorite analogs. Journal of Geophysical Research (Planets), 115:6005–.

Clark, B.E. and 14 co-authors, (2011). Asteroid (101955) 1999 RQ36: Spectroscopy from 0.4 to 2.4 µm and meteorite analogs. Icarus 216, 462-475.

Clark, R. (1979). Planetary reflectance measurements in the region of planetary thermal emission. Icarus, 40(1):94–103.

Clark, R. N. (1983). Spectral properties of mixtures of montmorillonite and dark carbon grains: Implications for remote sensing minerals containing chemically and physically adsorbed water. 88:10635–10644.

Clark, R. N. (1999). Spectroscopy of Rocks and Minerals, and Principles of Spectroscopy. Online at http://speclab.cr.usgs.gov/PAPERS.refl-mrs/refl4.html

Clark, R. N. (2009). Detection of Adsorbed Water and Hydroxyl on the Moon. Science, 326:562–.

Clark, R. N., King, T. V., Klejwa, M., Swayze, G. A., and Vergo, N. (1990). High spectral resolution reflectance spectroscopy of minerals. Journal of Geophysical Research: Solid Earth, 95(B8):12653–12680.

Clark, R. N., Swayze, G. A., Gallagher, A. J., King, T. V. V., Calvin, W. M. (1993). The U. S. Geological Survey, Digital Spectral Library: Version 1 (0.2 to 3.0µm). U. S. Geological Survey, Open File Report 93-592.

Clark, R.N., Swayze, G. A., Wise, R., Livo, K. E., Hoefen, T. M., Kokaly, R. F., and Sutley, S. J. (2007), USGS Digital Spectral Library splib06a, *U.S. Geological Survey, Data Series 231*.

Clark, R. N., Cruikshank, D. P., Jaumann, R., Brown, R. H., Stephan, K., Dalle Ore, C. M., Eric Livo, K., Pearson, N., Curchin, J. M., Hoefen, T. M., Buratti, B. J., Filacchione, G., Baines, K. H., and Nicholson, P. D. (2012). The surface composition of Iapetus: Mapping results from Cassini VIMS. Icarus, 218:831–860.

Cloutis, E. A., Hiroi, T., Gaffey, M. J., Alexander, C. M. O. ., and Mann, P. (2011a). Spectral reflectance properties of carbonaceous chondrites: 1. CI chondrites. Icarus, 212:180–209.

Cloutis, E. A., Hudon, P., Hiroi, T., Gaffey, M. J., and Mann, P. (2011b). Spectral reflectance properties of carbonaceous chondrites: 2. CM chondrites. Icarus, 216:309–346.

Cloutis, E., Hudon, P., Hiroi, T., and Gaffey, M. (2012). Spectral reflectance properties of carbonaceous chondrites 4: Aqueously altered and thermally metamorphosed meteorites. Icarus, 220(2):586–617.

Consolmagno, G. J., Weidenschilling, S. J., and Britt, D. T. (2003). Forming Well-compacted Meteorites by Shock Events in the Solar Nebula. Meteoritics and Planetary Science Supplement, 38:5247.

Cohen, M., Witteborn, F.C., Roush, T., Bregman, J., Wooden, D. (1998). Spectral irradiance calibration in the infrared. VIII. 5–14 micron spectroscopy of the Asteroids Ceres, Vesta, and Pallas. Astron. J. 115, 1671–1679

Cruikshank, D., Dalle Ore, C., Roush, T., Geballe, T., Owen, T., de Bergh, C., Cash, M., and Hartmann, W. (2001). Constraints on the composition of trojan asteroid 624 hektor. Icarus, 153(2):348–360.

Cuzzi, Jeffrey N. Robert C. Hogan, William F. Bottke (2010). Towards initial mass functions for asteroids and Kuiper Belt Objects Icarus 208:518–538

Clayton R. N. and Mayeda T. K. 1999. Oxygen isotope studies of carbonaceous chondrites. Geochimica et Cosmochimica Acta 63:2089–2104.

de León, J., Pinilla-Alonso, N., Campins, H., Licandro, J., & Marzo, G. A. (2012). Near-infrared spectroscopic survey of B-type asteroids: Compositional analysis, Icarus:218, 196-206.

De Luise, F.; Dotto, E.; Fornasier, S.; Barucci, M. A.; Pinilla-Alonso, N.; Perna, D.; Marzari, F. (2010). A peculiar family of Jupiter Trojans: The Eurybates, Icarus, 209; 586-590

Denevi, B. W.; Blewett, D. T.; Buczkowski, D. L.; Capaccioni, F.; Capria, M. T.; De Sanctis, M. C.; Garry, W. B.; Gaskell, R. W.; Le Corre, L.; Li, J.-Y.; (2012). Pitted Terrain on Vesta and Implications for the Presence of Volatiles, Science, Volume 338; 246

De Bergh, C., Boehnhardt, H., Barucci, M., Lazzarin, M., Fornasier, S., Romon-Martin, J., Tozzi, G., Doressoundiram, A., and Dotto, E. (2004). Aqueous altered silicates at the surface of two plutinos? Astronomy & Astrophysics, 416(2):791–798.

De Sanctis, M. C., Combe, J.-P., Ammannito, E., Palomba, E., Longobardo, A., McCord, T. B., Marchi, S., Capaccioni, F., Capria, M. T., Mittlefehldt, D. W., Pieters, C. M., Sunshine, J., Tosi, F., Zambon, F., Carraro, F., Fonte, S., Frigeri, A., Magni, G., Raymond, C. A., Russell, C. T., and Turrini, D. (2012). Detection of Widespread Hydrated Materials on Vesta by the VIR Imaging Spectrometer on board the Dawn Mission. Ap. J. Lett., 758:L36.

De Sanctis, M., Coradini, A., Ammannito, E., Filacchione, G., Capria, M., Fonte, S., Magni, G., Barbis, A., Bini, A., Dami, M., et al. (2012). The VIR spectrometer. In The Dawn Mission to Minor Planets 4 Vesta and 1 Ceres, pages 329–369. Springer. 42(3):501–513.

DeMeo, F. E.; Carry, B.(2014). Solar System evolution from compositional mapping of the asteroid belt. Nature 505:629-634.

DeMeo, F. E.; Carry, B.(2013). The taxonomic distribution of asteroids from multi-filter all-sky photometric surveys. Icarus, 226, Issue: 723-741.

DeMeo, F. E., Binzel, R. P., Slivan, S. M., and Bus, S. J. (2009). An extension of the Bus asteroid taxonomy into the near-infrared. Icarus, 202:160–180.

Dotto, E.; Fornasier, S.; Barucci, M. A.; Licandro, J.; Boehnhardt, H.; Hainaut, O.; Marzari, F.; de Bergh, C.; De Luise, F. (2006). The surface composition of Jupiter Trojans: Visible and near-infrared survey of dynamical families, Icarus, 183; 420-434

Dyar, M. D., Hibbitts, C. A., and Orlando, T. M. (2010). Mechanisms for incorporation of hydrogen in and on terrestrial planetary surfaces. Icarus, 208:425–437.

Emery, J. P. and Brown, R. H. (2003). Constraints on the surface composition of Trojan asteroids from near-infrared (0.8-4.0 μm) spectroscopy. Icarus, 164:104–121.

Emery, J.P. and R.H. Brown (2004). The surface composition of Trojan asteroids: Constraints set by scattering theory. *Icarus* 170, 131-152.

Emery, J.P., Cruikshank, D.P., Van Cleve, J. (2006). Thermal emission spectroscopy (5.2 – 38 μm) of three Trojan asteroids with the Spitzer Space Telescope: Detection of fine-grained silicates. Icarus 182, 496-512.

Farmer, V. (1974). The layer silicates. The infrared spectra of minerals, 4:331–363.

Feierberg, M. A., Lebofsky, L. A., and Larson, H. P. (1981). Spectroscopic evidence for aqueous alteration products on the surfaces of low-albedo asteroids. Geochim. et Cosmochim. Acta, 45:971–981.

Feldman, W.; Maurice, S.; Lawrence, D.; Little, R.; Lawson, S.; Gasnault, O.; Wiens, R.; Barraclough, B.; Elphic, R.; Prettyman, H.; (2001). Evidence for water ice near the lunar poles, Journal of Geophysical Research, 106; 23231-23252

Florczak, M., Lazzaro, D., Mothé-Diniz, T., Angeli, C. A., and Betzler, A. S. (1999). A spectroscopic study of the Themis family. Astron. Astroph. Suppl., 134:463–471.

Fornasier, S., Doressoundiram, A., Tozzi, G., Barucci, M., Boehnhardt, H., de Bergh, C., Delsanti, A., Davies, J., and Dotto, E. (2004). ESO large program on physical studies of trans-neptunian objects and centaurs: Final results of the visible spectrophotometric observations. Astronomy & Astrophysics, 421:353–363.

Fornasier, S., Barucci, M., De Bergh, C., Alvarez-Candal, A., DeMeo, F., Merlin, F., Perna, D., Guilbert, A., Delsanti, A., Dotto, E., et al. (2009). Visible spectroscopy of the new ESO large programme on trans-neptunian objects and centaurs: final results. Astronomy & Astrophysics, 508(1):457–465.

Fornasier, S.; Dotto, E.; Hainaut, O.; Marzari, F.; Boehnhardt, H.; De Luise, F.; Barucci, M. A.; (2007). Visible spectroscopic and photometric survey of Jupiter Trojans: Final results on dynamical families, Icarus, Volume 190, Issue 2, p. 622-642

Fornasier, S., Clark, B. E., and Dotto, E. (2011). Spectroscopic survey of X-type asteroids. Icarus, 214(1):131–146.

Fornasier, S.; Lantz, C.; Barucci, M. A.; Lazzarin, M. (2014). Aqueous alteration on main belt primitive asteroids: Results from visible spectroscopy, Icarus, 233; 163-178

Fraeman, A. A.; Murchie, S. L.; Arvidson, R. E.; Clark, R. N.; Morris, R. V.; Rivkin, A. S.; Spectral absorptions on Phobos and Deimos in the visible/near infrared wavelengths and their compositional constraints, Icarus, 229; 196-205

Gendrin, A., Langevin, Y., and Erard, S. (2005). ISM observation of phobos reinvestigated: Identification of a mixture of olivine and low-calcium pyroxene. 110. doi:10.1029/2004JE002245.

Gomes, R., Levison, H. F., Tsiganis, K., and Morbidelli, A. (2005). Origin of the cataclysmic Late Heavy Bombardment period of the terrestrial planets. Nature, 435:466–469.

Gounelle M (2011) The asteroid–comet continuum: In search of lost primitivity. Elements 7: 29-34.

Gradie, J., Tedesco, E., (1982). Compositional structure of the asteroid belt. Science 216 (June), 1405–1407.

Groussin, O.; Sunshine, J. M.; Feaga, L. M.; Jorda, L.; Thomas, P. C.; Li, J.-Y.; A'Hearn, M. F.; Belton, M. J. S.; Besse, S.; Carcich, B.; (2013). The temperature, thermal inertia, roughness and color of the nuclei of Comets 103P/Hartley 2 and 9P/Tempel 1, Icarus, 222; 580-594

Hanson, H. M., Howell, E., Magri, C., and Nolan, M. (2006). Correlating Arecibo radar and IRTF near-infrared spectral observations of 105 Artemis. In Bulletin of the American Astronomical Society, volume 38, page 933.

Harmon, J. K.; Slade, M. A.; Rice, M.S., 2011, Radar imagery of Mercury's putative polar ice: 1999-2005 Arecibo results, Icarus 211, pp. 37-50.

Harris, A. W. and Lagerros, J. S. V. (2002). Asteroids in the thermal infrared. In Bottke, W., Cellino, A., Paolicchi, P., and Binzel, R. P., editors, Asteroids III, pages 205–218. University of Arizona Press, Tucson.

Hasegawa, S., Murakawa, K., Ishiguro, M., Nonaka, H., Takato, N., Davis, C. J., Ueno, M., and Hiroi, T. (2003). Evidence of hydrated and/or hydroxylated minerals on the surface of asteroid 4 Vesta. GRL, 30:DOI 10.1029/2003GL01862.

Hiroi, T., Zolensky, M. E., Pieters, C. M., and Lipschutz, M. E. (1996). Thermal metamorphism of the C, G, B, and F asteroids seen from the 0.7 µm, 3 µm and UV absorption strengths in comparison with carbonaceous chondrites. 31:321–327.

Howell, E. S., Rivkin, A. S., Vilas, F., Magri, C., Nolan, M. C., Vervack, Jr., R. J., and Fernandez, Y. R. (2011). Hydrated silicates on main-belt asteroids: Correlation of the 0.7- and 3-micron absorption bands, EPSC-DPS Joint Meeting 2011:637.

Howell, E. S., Rivkin, A. S., and Vilas, F. (2005). Water Distribution in the Asteroid Belt. LPI Contributions, 1267:18.

Hsieh , H. H. and 41 co-authors (2012). Discovery of Main-belt Comet P/2006 VW$_{139}$ by Pan-STARRS1. ApJ:748, L15

Hsieh, H.H., Jewitt, D.C., Fernandez, Y.R. (2004) The Strange Case of 133P/Elst-Pizarro: A Comet among the Asteroids. The Astronomical Journal, Volume 127, Issue 5, pp. 2997-3017.

Hsieh, H. H. and Jewitt, D. (2006). A Population of Comets in the Main Asteroid Belt. Science, 312:561–563.

Hsieh, H. H., Yang, B., Haghighipour, N., et al. (2012) Discovery of Main-belt Comet P/2006 VW$_{139}$ by Pan-STARRS1, ApJ, 748L, 15H.

Hsieh, H., Jewitt, D., Lacerda, P., Lowry, S., Snodgrass, C. (2011a) The return of activity in main-belt comet 133P/Elst-Pizarro, Monthly Notices of the Royal Astronomical Society, 403: 363-377.

Hsieh, H., Meech, K., Pittichova, J. (2011b) Main-belt Comet 238P/Read Revisited, Ap. J. lett. 736:L18

Hunt, G. R. (1977). Spectral signatures of particulate minerals in the visible and near infrared. Geophysics,

Iwata, T., Kitazato, K., Abe, M., Ohtake, M., Matsuura, S., Tsumura, K., Hirata, N., Honda, C., Takagi, Y., Nakauchi, Y., Hiroi, T., Senshu, H., Arai, T., Nakamura, T., Matsunaga, T., Komatsu, M., Takato, N., and Watanabe, S. (2013). Results of the Critical Design for NIRS3: The Near Infrared Spectrometer on Hayabusa-2. In Lunar and Planetary Science Conference, volume 44 of Lunar and Planetary Science Conference, page 1908.

Jarvis, K. S., Vilas, F., Larson, S. M., and Gaffey, M. J. (2000). JVI Himalia: New compositional evidence and interpretations for the origin of Jupiter's small satellites. Icarus, 145:445–453.

Jewitt, D. (2012). The active asteroids. The Astronomical Journal, 143(3):66.

Jewitt, D., Li, J., and Agarwal, J. (2013). The dust tail of asteroid (3200) Phaethon. The Astrophysical Journal Letters, 771(2):L36.

Jewitt, D., Agarwal, J., Weaver, H., Mutchler, M., and Larson, S., (2015). Episodic Ejection from Active Asteroid 311P/PANSTARRS. The Astrophysical Journal, 798, id 109.

Jones, T. D. (1989). An infrared reflectance study of water in outer belt asteroids: Clues to composition and origin. Ph.D. dissertation, University of Arizona, Tucson, AZ.

Jones, T. D., Lebofsky, L. A., Lewis, J. S., and Marley, M. S. (1990). The composition and origin of the C, P, and D asteroids: Water as a tracer of thermal evolution in the outer belt. Icarus, 88:172–192

King, T. V. V. and Clark, R. N. (1997). The presence of a single absorption feature: What it does and doesn't imply. LPSC, 28:727–728.

King, T. V. V., Clark, R. N., Calvin, W. M., Sherman, D. M., and Brown, R. H. (1992). Evidence for ammonium-bearing minerals on Ceres. Science, 255:1551–1553.

Küppers, M.; O'Rourke, L.; Bockelée-Morvan, D.; Zakharov, V.; Lee, S.; von Allmen, P.; Carry, B.; Teyssier, D.; Marston, A.; Müller, T.; (2014). Localized sources of water vapour on the dwarf planet (1)Ceres, Nature, 505: 525-527

Landsman, Z. A., Campins, H., Hargrove, K., Pinilla-Alonso, N., Emery, J., and Ziffer, J. (2013). An investigation of the 3-µm feature in M-type asteroids. In AAS/Division for Planetary Sciences Meeting Abstracts, volume 45.

Lauretta, D.S. and 28 co-authors 2014. The OSIRIS-REx target asteroid 101955 Bennu: Constraints on its physical, geological, and dynamical nature from astronomical observations. *Meteor. Planet. Sci.* in press.

Lazzaro, D., Angeli, C. A., Carvano, J. M., Mothé-Diniz, T., Duffard, R., and Florczak, M. (2004). S3OS2: the visible spectroscopic survey of 820 asteroids. Icarus, 172:179–220.

Lebofsky, L. A. (1980). Infrared reflectance spectra of asteroids: A search for water of hydration. 85:573–585.

Lebofsky, L. A. (1980). Infrared reflectance spectra of asteroids: A search for water of hydration. Astron. J., 85:573–585.

Lebofsky, L. A. and Spencer, J. R. (1989). Radiometry and thermal modeling of asteroids. In Binzel, R., Gehrels, T., and Matthews, M. S., editors, Asteroids II, pages 128–147. University of Arizona Press, Tucson.

Lebofsky, L. A., Jones, T. D., Owensby, P. D., Feierberg, M. A., and Consolmagno, G. J. (1990). The nature of low albedo asteroids from 3-µm spectrophotometry. Icarus, 83:12–26.

Lee M. (1993). The petrography, mineralogy and origins of calcium sulphate within the Cold Bokkeveld CM carbonaceous chondrite. Meteoritics 28:53–62.

Levison, H.F., Bottke, W.F., Gounelle, M., Morbidelli, A., Nesvorny, D., Tsiganis, K. (2009) Contamination of the asteroid belt by primordial trans-Neptunian Objects. Nature, 460, 364-366.

Licandro, J.; Campins, H.; Mothé-Diniz, T.; Pinilla-Alonso, N.; de León, J. (2007). The nature of comet-asteroid transition object (3200) Phaethon, Astronomy and Astrophysics, Volume: 751-757

Licandro, J., Campins, H., Kelley, M., Hargrove, K., Pinilla-Alonso, N., Cruikshank, D., Rivkin, A. S., and Emery, J. (2011a). (65) Cybele: detection of small silicate grains, water-ice, and organics. A&A, 525:A34+.

Licandro, J.; Campins, H.; Tozzi, G. P.; de León, J.; Pinilla-Alonso, N.; Boehnhardt, H.; Hainaut, O. R. Testing the comet nature of main belt comets. The spectra of 133P/Elst-Pizarro and 176P/LINEAR (2011b) A&A...532A..65L

Licandro, J., de León, J., Kelley, M. S., Emery, J., Rivkin, A., Pinilla-Alonso, N., Mothè-Diniz, T., Campins, H., Alí-Lagoa, V. (2011c), Multi-wavelength study of activated asteroid (596) Scheila, EPSC Abstracts Vol. 6, EPSC-DPS2011-1109-1

Licandro, J., Moreno, F., de León, J., Tozzi, G. P., Lara, L. M., Cabrera-Lavers, A. (2013). Exploring the nature of new main-belt comets with the 10.4 m GTC telescope: (300163) 2006 VW139, A&A 550: 7

Licandro, J, Alvarez-Iglesias, C., Cabrera-Lavers, A., Alí-Lagoa, V., Pinilla-Alonso, N., Campins, H., de León, J., and Kelley, M. (2014), The GTC mid-infrared spectroscopic program of primitive outer-belt asteroids, In Proceedings of Asteroids, Comets, Meteors 2014, Muinonen et al., Eds.

Lunine, J. and Reid, I. (2006). Astrobiology: a multidisciplinary approach. Physics Today. 59: 58

Mainzer, A., Grav, T., Masiero, J., Bauer, J., Wright, E., Cutri, R., Walker, R., and McMillan, R. (2011). Thermal model calibration for minor planets observed with wise/neowise: Comparison with infrared astronomical satellite. The Astrophysical Journal Letters, 737(1):L9.

Marchi, S., Delbo', M., Morbidelli, A., Paolicchi, P., and Lazzarin, M. (2009). Heating of near-Earth objects and meteoroids due to close approaches to the Sun. Mon. Not. Royal Ast. Soc., 400:147–153.

Martin R. and Livio M. (2012). On the evolution of the snow line in protoplanetary discs. Monthly Notices of the Royal Astronomical Society: Letters, 425: L6-L9

Mastrapa, R. M.; Sandford, S. A.; Roush, T. L.; Cruikshank, D. P.; Dalle Ore, C. M, (2009). Optical Constants of Amorphous and Crystalline $H_2O$-ice: 2.5-22 μm (4000-455 $cm^{-1}$) Optical Constants of $H_2O$-ice, The Astrophysical Journal, 701: 1347-1356

McAdam, M., Sunshine, J.M., Kelley, M.S. 2013. Composition and degree of alteration of dark asteroids. $45^{th}$ AAS/Division for Planetary Sciences, abstract #205.10.

McCord, T. B.; Li, J.-Y.; Combe, J.-P.; McSween, H. Y.; Jaumann, R.; Reddy, V.; Tosi, F.; Williams, D. A.; Blewett, D. T.; Turrini, D. (2012). Dark material on Vesta from the infall of carbonaceous volatile-rich material, Nature, 491: 83-86

McSween H. Y. 1979. Alteration in CM carbonaceous chondrites inferred from modal and chemical variations in matrix. Geochimica et Cosmochimica Acta 43:1761-1770

Milliken, R. E. and Rivkin, A. S. (2009). Brucite and carbonate assemblages from altered olivine-rich materials on Ceres. Nature Geoscience, 2:258–261.

Milliken, R. E., Mustard, J. F., Poulet, F., Jouglet, D., Bibring, J.-P., Gondet, B., and Langevin, Y. (2007). Hydration state of the martian surface as seen by mars express omega: 2. $H_2O$ content of the surface. Journal of Geophysical Research: Planets (1991–2012), 112(E8).

Mommert, M.; Farnocchia, D.; Hora, J. L.; Chesley, S. R.; Trilling, D. E.; Chodas, P. W.; Mueller, M.; Harris, A. W.; Smith, H. A.; Fazio, G. G., (2014) The Discovery of Cometary Activity in Near-Earth Asteroid (3552) Don Quixote by Mommert et al., Astron J. 781, id. 25

Morbidelli, A., Levison, H. F., Tsiganis, K., and Gomes, R. (2005). Chaotic capture of Jupiter's Trojan asteroids in the early Solar System. Nature, 435:462–465.

Morbidelli, A., Bottke, W. F., Nesvorny`, D., and Levison, H. F. (2009). Asteroids were born big. Icarus, 204(2):558–573.

Morbidelli, A., Brasser, R., Gomes, R., Levison, H. F. & Tsiganis, K.(2010). Evidence from the asteroid belt for a violent past evolution of Jupiter's orbit. Astron. J. 140, 1391–1401.

Moreno, F., Licandro, J., Ortiz, J., Lara, L. M., Alí-Lagoa, V., Vaduvescu, O., Morales, N., Molina, A., and Lin, Z.-Y. (2011). (596) Scheila in outburst: A probable collision event in the main asteroid belt. The Astrophysical Journal, 738(2):130.

Moreno, F.; Cabrera-Lavers, A.; Vaduvescu, O.; Licandro, J.; Pozuelos, F. (2013). The Dust Environment of Main-Belt Comet P/2012 T1 (PANSTARRS), The Astrophysical Journal Letters, 770:L30

Murchie, S. and Erard, S. (1996). Spectral properties and heterogeneity of Phobos from measurements by Phobos 2. Icarus, 123:63–86.

Nedelcu, D. A., Birlan, M., Vernazza, P., Descamps, P., Binzel, R. P., Colas, F., Kryszczynska, A., and Bus, S. J. (2007). Near infra-red spectroscopy of the asteroid 21 Lutetia. II. Rotationally resolved spectroscopy of the surface. A&A, 470:1157–1164.

Nesvorný, D; Bottke, W.; Vokrouhlický, D.; Sykes, M.; Lien, D.; Stansberry, J.; (2008). Origin of the Near-Ecliptic Circumsolar Dust Band, he Astrophysical Journal, 679: L143-L146

Noble, S. K., Pieters, C. ., Taylor, L. A., Morris, R. V., Allen, C. C., McKay, D. S., and Keller, L. P. (2001). The optical properties of the finest fraction of lunar soil: Implications for space weathering. 36:31–42.

Okamura, N., Hasegawa, S., Usui, F., Hiroi, T., Ootsubo, T., Müller, T. G., and Sugita, S. (2014a). Spec- troscopic Observations of Dark Main-Belt Asteroids in the 2.5-3.1 μm Range. In Lunar and Planetary Science Conference, volume 45 of Lunar and Planetary Inst. Technical Report, page 1375.

Okamura, N., Sugita, S., Kamata, S., Usui, F., Hiroi, T., Ootsubo, T., Müller, T. G., Sakon, I., and Hasegawa, S. (2014b). Principal-Component Analysis of the Continuous 3-μm Spectra of Low-Albedo Asteroids Observed with the AKARI Satellite. In Lunar and Planetary Science Conference, volume 45 of Lunar and Planetary Inst. Technical Report, page 2446.

Osawa T., Kagi H., Nakamura T., and Noguchi T. (2005). Infrared spectroscopic taxonomy for carbonaceous chondrites from speciation of hydrous components. Meteoritics & Planetary Science 40:71–86.

Pieters, C. M., Goswami, J. N., Clark, R. N., Annadurai, M., Boardman, J., Buratti, B., Combe, J., Dyar, M. D., Green, R., Head, J. W., Hibbitts, C., Hicks, M., Isaacson, P., Klima, R., Kramer, G., Kumar, S., Livo, E., Lundeen, S., Malaret, E., McCord, T., Mustard, J., Nettles, J., Petro, N., Runyon, C., Staid, M., Sunshine, J., Taylor, L. A., Tompkins, S., and Varanasi, P. (2009). Character and Spatial Distribution of OH/H2O on the Surface of the Moon Seen by $M^3$ on Chandrayaan-1. Science, 326:568–.

Pinilla-Alonso, N., 2011. Ice vs. Goethite on Themis. Workshop on Water in Asteroids and Meteorites. Observatoire de Paris, Paris, France.

Prettyman, T.; Mittlefehldt, D.; Yamashita, N.; Lawrence, D..; Beck, A.; Feldman, W.; McCoy, T.; McSween, H.; Toplis, M.; Titus, T. (2012). Elemental Mapping by Dawn Reveals Exogenic H in Vesta's Regolith, Science, Volume 338; 242

Reddy, Vishnu; Le Corre, Lucille; O'Brien, David P.; Nathues, Andreas; Cloutis, Edward A.; Durda, D.; Bottke, W.; Bhatt, M.; Nesvorny, D.; Buczkowski, D.; (2012). Delivery of dark material to Vesta via carbonaceous chondritic impacts, Icarus, 221; 544-559

Rivkin, A. (2012). The fraction of hydrated C-complex asteroids in the asteroid belt from SDSS data. Icarus.

Rivkin, A. S., Lebofsky, L. A., Clark, B. E., Howell, E. S., and Britt, D. T. (2000). The nature of M-class asteroids in the 3-µm region. Icarus, 145:351–368.

Rivkin, A. S., Brown, R. H., Trilling, D. E., Bell, J. F., I., and Plassmann, J. H. (2002a). Infrared spectrophotometry of Phobos and Deimos. Icarus, 156:64–75.

Rivkin, A. S., Howell, E. S., Vilas, F., and Lebofsky, L. A. (2002b). Hydrated minerals on asteroids: The astronomical record. In Bottke, W., Cellino, A., Paolicchi, P., and Binzel, R. P., editors, Asteroids III, pages 235–253. University of Arizona Press, Tucson.

Rivkin, A. S., Davies, J. K., Johnson, J. R., Ellison, S. L., Trilling, D. E., Brown, R. H., and Lebofsky, L. A. (2003). Hydrogen concentrations on C-class asteroids derived from remote sensing. 38:1383–1398.

Rivkin, A. S., McFadden, L. A., Binzel, R. P., and Sykes, M. (2006). Rotationally-resolved spectroscopy of Vesta I: 2 4 µm region. Icarus, 180:464–472.

Rivkin, A.S.; Volquardsen, E.L.(2010). Rotationally-resolved spectra of Ceres in the 3-µm region. Icarus, Volume 206, Issue 1, p. 327-333.

Rivkin, A. S. and Emery, J. P. (2010). Detection of ice and organics on an asteroidal surface. Nature, 464:1322–1323.

Rivkin, A. S., Clark, B. E., Ockert-Bell, M. E., Shepard, M. K., Volquardsen, E. L., Howell, E. S., and Bus, S. J. (2011). Asteroid 21 Lutetia at 3 µm: Observations with IRTF SpeX. Icarus, 216:62–68.

Rivkin, A. S., Howell, E. S., Emery, J. P., Volquardsen, E. L., and DeMeo, F. E. (2012). Toward a taxonomy of asteroid spectra in the 3-µm region. In European Planetary Science Congress 2012, page 359.

Rivkin, A. S., Howell, E. S., Vervack Jr, R. J., Magri, C., Nolan, M. C., Fernandez, Y. R., Cheng, A. F., Antonietta Barucci, M., and Michel, P. (2013a). The NEO (175706) 1996 FG3 in the 2–4µm spectral region: Evidence for an aqueously altered surface. Icarus, 223:493–498.

Rivkin, A. S., Howell, E., Emery, J., and Volquardsen, E. (2013b). The Ch asteroids: Connecting a visible taxonomic class to a 3-µm spectral shape. In AAS/Division for Planetary Sciences Meeting Abstracts, volume 45.

Rivkin, A. S., Howell, E. S., Emery, J. P., and Sunshine, J. M. (2013c). Does the Solar Wind Create OH on NEO Surfaces?: Observations of 433 Eros and 1036 Ganymed. In Lunar and Planetary Science Conference, volume 44 of Lunar and Planetary Inst. Technical Report, page 2070.

Rivkin, A. S., Asphaug, E., and Bottke, W. F. (2014a). The Case of the Missing Ceres Family. Icarus 243, 429-439.

Rivkin, A. S., Howell, E., and Emery, J. (2014b). The LXD-mode Main-belt/NEO Observing Program (LMNOP): Results. In Proceedings of Asteroids, Comets, Meteors 2014, Muinonen et al., Eds.

Rubin, A. E., Trigo-Rodríguez, J. M., Huber, H., and Wasson, J. T. (2007). Progressive aqueous alteration of CM carbonaceous chondrites. Geochimica et Cosmochimica Acta, 71(9):2361–2382.

Ryskin, Y. I. (1974). The vibrations of protons in minerals: hydroxyl, water and ammonium. The infrared spectra of minerals, pages 137–181.

Salisbury, J. W., D'Aria, D. M., and Jarosewich, E. (1991). Midinfrared (2.5–13.5 μm) reflectance spectra of powdered stony meteorites. Icarus, 92:280–297.

Sato, K., Miyamoto, M., and Zolensky, M. E. (1997). Absorption bands near three micrometers in diffuse reflectance spectra of carbonaceous chondrites: Comparison with asteroids. 32:503–507.

Sawyer, S. R. (1991). A High resolution CCD spectroscopic survey of low-albedo main belt asteroids. Ph.D. dissertation, University of Texas, Austin, TX.

Schorghofer, N. (2008). The Lifetime of Ice on Main Belt Asteroids. Ap. J., 682:697–705.

Seu, R., Biccari, D., Orosei, R., Lorenzoni, L., Phillips, R., Marinangeli, L., Picardi, G., Masdea, A., and Zampolini, E. (2004). SHARAD: The MRO 2005 shallow radar. Planetary and Space Science, 52(1):157–166.

Shands, A. L. (1949) Mean Precipitable Water in the United States. U. S. Department of Commerce Division of Climatological and Hydrologic Services, Technical Paper No. 10.

Shepard, M. K., Taylor, P. A., Nolan, M. C., Howell, E. S., Springmann, A., Giorgini, J. D., Warner, B. D., Harris, A. W., Stephens, R., Merline, W. J., et al. (2015). A radar survey of M-and X-class asteroids. III. insights into their composition, hydration state, & structure. Icarus, 245:38–55.

Starukhina, L. (2001). Water detection on atmosphereless celestial bodies: Alternative explanations of the observations. 106:14701–14710.

Stern, S. A., Slater, D. C., Scherrer, J., Stone, J., Dirks, G., Versteeg, M., Davis, M., Gladstone, G. R., Parker, J. W., Young, L. A., et al. (2008). Alice: The ultraviolet imaging spectrograph aboard the New Horizons Pluto–Kuiper belt mission. Space Science Reviews, 140(1-4):155–187.

Stern, S. A.; Parker, J. Wm.; Feldman, P. D.; Weaver, H. A.; Steffl, A.; A'Hearn, M. F.; Feaga, L.; Birath, E.; Graps, A.; Bertaux, J.-L.; (2011). Ultraviolet Discoveries at Asteroid (21) Lutetia by the Rosetta Alice Ultraviolet Spectrograph, The Astronomical Journal, 141; 3

Stewart, B. D.; Pierazzo, E.; Goldstein, D. B.; Varghese, P. L.; Trafton, L.M. (2011). Simulations of a comet impact on the Moon and associated ice deposition in polar cold traps. Icarus 216, 1-16.

Su, K. Y., Rieke, G. H., Malhotra, R., Stapelfeldt, K. R., Hughes, A. M., Bonsor, A., Wilner, D. J., Balog, Z., Watson, D. M., Werner, M. W., et al. (2013). Asteroid belts in debris disk twins: Vega and Fomalhaut. The Astrophysical Journal, 763(2):118.

Sunshine, J. M., Farnham, T. L., Feaga, L. M., Groussin, O., Merlin, F., Milliken, R. E., and A'Hearn, M. F. (2009). Temporal and Spatial Variability of Lunar Hydration As Observed by the Deep Impact Spacecraft. Science, 326:565–.

Takir, D. and Emery, J. (2012). Outer main belt asteroids: Identification and distribution of four 3-μm spectral groups. Icarus.

Takir, D., Emery, J. P., Mcsween, H. Y., Hibbitts, C. A., Clark, R. N., Pearson, N., and Wang, A. (2013). Nature and degree of aqueous alteration in CM and CI carbonaceous chondrites. Meteoritics & Planetary Science, 48(9):1618–1637.

Takir, D., Emery, J., and McSween, H. (2015). Toward an understanding of phyllosilicate mineralogy in the outer main belt region. Icarus, in press.

Tsiganis, K., Gomes, R., Morbidelli, A., Levison H., (2005). Origin of the orbital architecture of the giant planets of the Solar System. Nature, 435:459-461

Vernazza, P.; Fulvio, D.; Brunetto, R.; Emery, J. P.; Dukes, C. A.; Cipriani, F.; Witasse, O.; Schaible, M. J.; Zanda, B.; Strazzulla, G.; Baragiola, R. A.; (2013), Paucity of Tagish Lake-like parent bodies in the Asteroid Belt and among Jupiter Trojans, Icarus, 225; 517-525

Vilas, F. (1994). A cheaper, faster, better way to detect water of hydration on solar system bodies. Icarus, 111:456–467.

Vilas, F. (1995). Is the U-B color sufficient for identifying water of hydration on solar system bodies? Icarus, 115:217–218.

Vilas, F. (2010). Composition of Jovian Outer Irregular Satellites from Reflectance Spectrophotometry: New MMT Data. BAAS 42: 951.

Vilas, F. (2013). Reflectance Spectrophotometry of the Irregular Outer Jovian Satellites as Insight to Solar System History. LPSC 44: 2900.

Vilas, F. and Gaffey, M. J. (1989). Phyllosilicate absorption features in main-belt and outer-belt asteroid reflectance spectra. Science, 246:790–792.

Vilas, F., Hatch, E. C., Larson, S. M., Sawyer, S. R., and Gaffey, M. J. (1993). Ferric iron in primitive asteroids - A 0.43-micron absorption feature. Icarus, 102:225–231.

Vilas, F., Larson, S. M., Stockstill, K. R., and Gaffey, M. J. (1996). Unraveling the Zebra: Clues to the Iapetus Dark Material Composition. Icarus, 124:262–267.

Vilas, F., Lederer, S. M., Gill, S. L., Jarvis, K. S., and Thomas-Osip, J. E. (2006). Aqueous alteration affecting the irregular outer planets satellites: Evidence from spectral reflectance. Icarus, 180:453–463.

Volquardsen, E. L., Rivkin, A. S., and Bus, S. J. (2007). Composition of hydrated near-Earth object (100085) 1992 UY4. Icarus, 187:464–468.

Vilas, F., Sollitt, L., and Sykes, M. V. (2013). Progress toward the operational Atsa Suborbital Observatory. Next Generation Suborbital Researcher's Conference 2013 abstracts.

Walsh, K. J., Morbidelli, A., Raymond, S. N., O'Brien, D. P., and Mandell, A. M. (2011). A low mass for Mars from Jupiter's early gas-driven migration. Nature, 475:206–209.

Watson, K., Murray, B., and Brown, H. (1961). On the possible presence of ice on the moon. Journal of Geophysical Research, 66(5):1598–1600.

Weidenschilling, S. (2011). Initial sizes of planetesimals and accretion of the asteroids. Icarus, 214(2):671–684.

Wigton, N., Emery, J. P., Rivkin, A. S., and Thomas, C. A. (2014). Near-Infrared (2 - 4 μm) spectroscopy of Near-Earth Asteroids: Searching for OH/H2O on small planetary bodies. In AAS/Division for Planetary Sciences Meeting Abstracts, volume 46 of AAS/Division for Planetary Sciences Meeting Abstracts, abs. no. 213.16.

Yang, B., Jewitt, D., and Bus, S. J. (2009). Comet 17P/Holmes in outburst: the near infrared spectrum. The Astronomical Journal, 137(5):4538.

Young, E. F., Osterman, S., Woodruff, R., Germann, L., Diller, J., Dinkel, K.,Dischner, Z., and Truesdale, N. (2013). Beyond HST: High Acuity Imaging from the Earth's Stratosphere. Next Generation Suborbital Researcher's Conference 2013 abstracts..

Zellner, B; Tholen, D. J.; Tedesco, E. F. (1985). The eight-color asteroid survey - Results for 589 minor planets. Icarus, 61:335-416.

Zolensky, M. E., Barrett, R., and Browning, L. (1993). Mineralogy and composition of matrix and chondrule rims in carbonaceous chondrites. 57:3123–3148.

Zolensky, M.E., Nakamura, K., Gounelle, M., Mikouchi,T, Kasama, T., Tachikawa, O., Tonui, E. (2002). Mineralogy of Tagish Lake: An ungropued type-2 carbonaceous chondrite. *Meteor. Planet. Sci*. 37, 737-761.

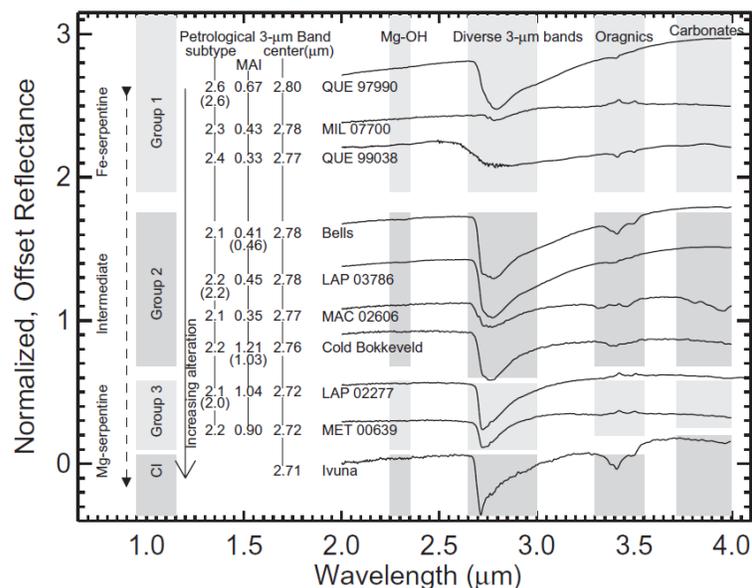

Figure 1. IR reflectance spectra of CM and CI carbonaceous chondrites measured under dry and vacuum conditions. The MAI and the petrologic subtype were determined applying the alteration scales of *Browning et al.* (1996) and *Rubin et al.* (2007), respectively. Our spectral investigation revealed distinct groups among CM chondrites: group 1 (QUE 97990, QUE 99038, and MIL 07700), group 2 (Bells, LAP 03786, MAC 02606, and Cold Bokkeveld), and group 3 (LAP 02277 and MET 00639). Ivuna is the only CI chondrite analyzed. The 3 μm band center decreases with increasing alteration. Adopted from *Takir et al.* (2013).

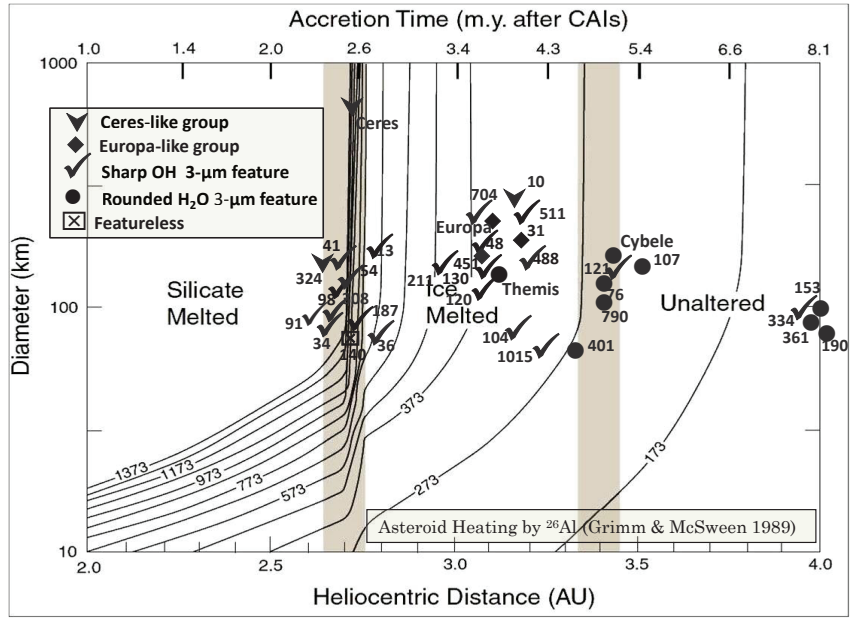

Figure 2. Asteroids classified according to the Takir 3-µm classes are plotted in the context of the thermal model of *Grimm and McSween* (1993). Adopted from *Takir and Emery* (2012).

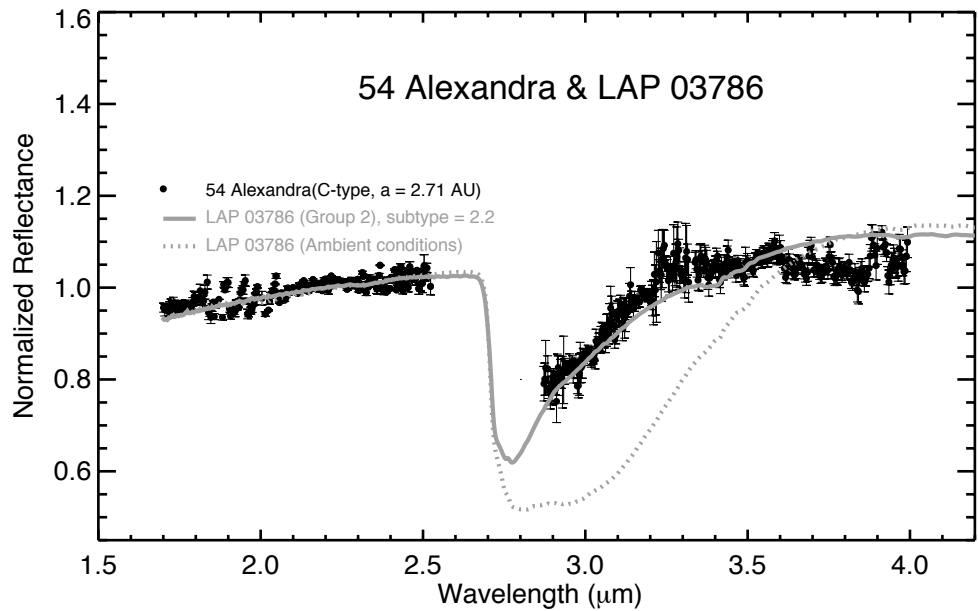

Figure 3. Meteorite LAP 03786 and 54 Alexandra, showing an example of the comparison between CM and CI chondrites and asteroids with the sharp/Pallas-type 3-μm feature. While asteroid spectra have been poor matches for meteorite spectra taken in ambient conditions (dotted line), meteorite spectra with terrestrial water removed (solid line) are much better matches. Figure from *Takir et al.* (2015).

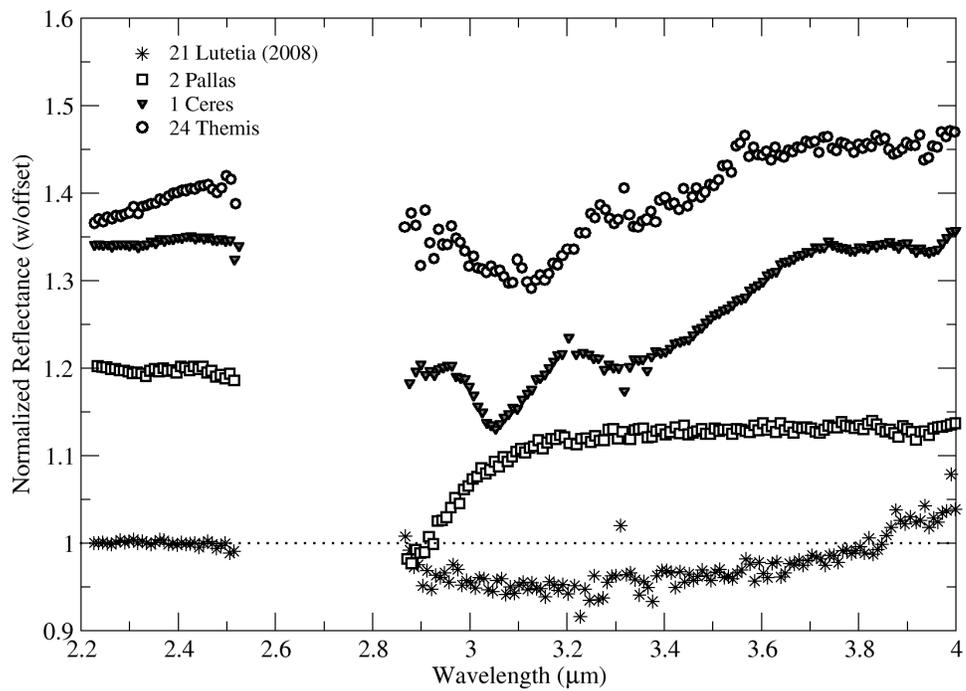

Figure 4: The diversity of asteroids in the 3-μm spectral region can be seen by comparing several of the largest objects. Each of these bodies shows evidence of different mineralogies, though the specific minerals in each case may still be under debate. Figure from Rivkin et al. (2011)

| Absorption Band Positions | | Composition | Notes |
|---|---|---|---|
| ~0.6-0.67 µm | Saponite | Caused by $Fe^{3+}$-$Fe^{2+}$ charge transfer | |
| ~0.7-0.75 µm | Serpentine | | |
| ~2.7-2.8 µm | Hydroxyl (OH) | Position can vary from 2.67-2.94 µm in phyllosilicates | |
| ~2.95 µm | Water in minerals | Position can vary | |
| ~3.1-3.2 µm | Water ice | Positions vary depending on crystalline vs. amorphous nature, temperature | |
| ~10-12.5 µm | Aluminum, magnesium, iron hydroxides | $Mg_2OH$ center wavelengths 15-17 µm | |

Table 1: Characteristics of the main absorption bands discussed in this chapter. Specific positions depend upon many factors, including temperature and specific mineral composition. Information taken from Clark et al. (1990), Cloutis et al. (2011a, 2011b), and Dyar et al. (2010), and references therein.

|  |  | All Asteroids | | C-complex asteroids | |
| --- | --- | --- | --- | --- | --- |
|  |  | 3-μm band absent | 3-μm band present | 3-μm band absent | 3-μm band present |
| 0.7-μm band present | Number of objects *Percent of objects* | 0 *0%* | 56 *35%* | 0 *0%* | 44 *63%* |
| 0.7-μm band absent | Number of objects *Percent of objects* | 57 *36%* | 47 *29%* | 9 *13%* | 17 *24%* |

Table 2: Correlation between the 0.7- and 3-μm bands in the total sample of available main-belt asteroid spectra and the sample of C-complex asteroids. A total of 160 main-belt asteroids have been observed at both 0.7 and 3 microns to investigate the hydration state. The bands are correlated for 113 objects, or 70% of the sample. The remaining objects have a 3 micron band, but no 0.7 micron band. No objects are found to have a 0.7 micron band, and lacking a 3 micron band, making the presence of the 0.7 band a good proxy for hydration, but only a lower limit on the number of hydrated objects. The correlation for 70 C-complex objects is shown in the right two columns. In cases where insufficient data exists to determine the Bus-DeMeo taxonomic class, we include Tholen C, B, F, and G taxonomic classes as well as CP and PC, but not those classified as P. The data sources from this table are Zellner et al. (1985), Lazzaro et al. (2004), Sawyer (1991), Vilas (1994), Lebofsky (1980), Lebofsky et al. (1990), Feierberg et al. (1985), Rivkin et al. (2000, 2002, 2012), Fornasier et al. (2011, 2014), and Howell et al. (2011).

| Example body | Takir class | Rivkin class | Interpretation |
| --- | --- | --- | --- |
| 511 Davida | Sharp | Pallas | CM-like phyllosilicates |
| 24 Themis | Rounded | Themis | Water ice frost, organics |
| 52 Europa | Europa | Themis | Water ice frost, organics |
| 1 Ceres | Ceres | Ceres | Brucite, carbonates |
| 21 Lutetia | -- | Lutetia | Goethite? |

Table 3: While a formal taxonomy has not yet been developed for objects in the 3-µm spectral region, the major groupings in such a taxonomy are already apparent. Presented here are the classes as defined by Takir et al. (2012) and Rivkin et al. (2012), with type examples and possible compositions for each group (also from these cited works).